%% file: EOMbbs.tex
\input harvmacM.tex
\input epsf

\preprint AEI--2014--011 DAMTP--2014--25

\title Multiparticle SYM equations of motion and pure spinor BRST blocks

\author Carlos R. Mafra\email{\dagger}{c.r.mafra@damtp.cam.ac.uk} and
	Oliver Schlotterer\email{\ddagger}{olivers@aei.mpg.de}

\address
$^\dagger$DAMTP, University of Cambridge
Wilberforce Road, Cambridge, CB3 0WA, UK

\address
$^\ddagger$Max--Planck--Institut f\"ur Gravitationsphysik
Albert--Einstein--Institut, 14476 Potsdam, Germany

\abstract

In this paper a multiparticle generalization of linearized ten-dimensional super Yang--Mills superfields is
proposed. Their equations of motions are shown to take the same form as in the single-particle case,
supplemented by contact terms. A recursive construction of these superfields is inspired by the
iterated OPEs among massless
vertex operators in the pure spinor formalism. An enlarged set of BRST-covariant pure spinor blocks is then defined in a
streamlined fashion and combined to  multiparticle vertex operators.
The latter can be used to  universally describe
tree-level subdiagrams in the perturbative open and closed superstring, regardless of the loop order. The inherent
symmetries of the multiparticle superfields are reproduced by structure constants of the gauge group, hinting a
natural appearance of the BCJ-duality between color and kinematics in the fundamentals of super Yang--Mills theory.
We present one-loop
applications where known scalar cohomology objects are systematically recomputed and a novel vector cohomology
particularly relevant to the closed string is constructed for arbitrary multiplicity.

\Date {April 2014}

\newif\iffig
\figfalse


\lref\WLA{
  N.~Berkovits,
  ``Infinite Tension Limit of the Pure Spinor Superstring,''
JHEP {\bf 1403}, 017 (2014).
[arXiv:1311.4156 [hep-th], arXiv:1311.4156].
}

\lref\WLB{
	H.~Gomez and E.~Y.~Yuan,
	``N-point tree-level scattering amplitude in the new Berkovits` string,''
	JHEP {\bf 1404}, 046 (2014).
	[arXiv:1312.5485 [hep-th]].
}

\lref\mathA{
G.~Melan\c con and C.~Reutenauer, ``Free Lie superalgebras, trees and chains
of partitions'', J. Algebraic Combin. 5 (1996), no. 4, 337--351
}

\lref\loday{
J.-L~Loday and M.~O.~Ronco, ``Hopf algebra of the planar binary trees'', Adv. Math. 139 (1998),
no. 2, 293--309.
}

\lref\BarreiroDPA{
  L.~A.~Barreiro and R.~Medina,
  ``RNS derivation of N-point disk amplitudes from the revisited S-matrix approach,''
[arXiv:1310.5942 [hep-th]].
}

\lref\KawaiXQ{
  H.~Kawai, D.~C.~Lewellen and S.~H.~H.~Tye,
  ``A Relation Between Tree Amplitudes of Closed and Open Strings,''
Nucl.\ Phys.\ B {\bf 269}, 1 (1986).
}

\lref\BjerrumBohrRD{
  N.~E.~J.~Bjerrum-Bohr, P.~H.~Damgaard and P.~Vanhove,
  ``Minimal Basis for Gauge Theory Amplitudes,''
Phys.\ Rev.\ Lett.\  {\bf 103}, 161602 (2009).
[arXiv:0907.1425 [hep-th]].
}

\lref\StiebergerHQ{
  S.~Stieberger,
  ``Open \& Closed vs. Pure Open String Disk Amplitudes,''
[arXiv:0907.2211 [hep-th]].
}

\lref\BerendsME{
  F.~A.~Berends and W.~T.~Giele,
  ``Recursive Calculations for Processes with n Gluons,''
Nucl.\ Phys.\ B {\bf 306}, 759 (1988)..
}

\lref\oneloopbb{
	C.R.~Mafra and O.~Schlotterer,
	``The Structure of n-Point One-Loop Open Superstring Amplitudes,''
	[arXiv:1203.6215 [hep-th]].
}
\lref\fivetree{
	C.R.~Mafra,
	``Simplifying the Tree-level Superstring Massless Five-point Amplitude,''
	JHEP {\bf 1001}, 007 (2010).
	[arXiv:0909.5206 [hep-th]].
}
\lref\nptMethod{
	C.R.~Mafra, O.~Schlotterer, S.~Stieberger and D.~Tsimpis,
	``A recursive method for SYM n-point tree amplitudes,''
	Phys.\ Rev.\ D {\bf 83}, 126012 (2011).
	[arXiv:1012.3981 [hep-th]].
}
\lref\nptTree{
	C.R.~Mafra, O.~Schlotterer and S.~Stieberger,
	``Complete N-Point Superstring Disk Amplitude I. Pure Spinor Computation,''
	Nucl.\ Phys.\ B {\bf 873}, 419 (2013).
	[arXiv:1106.2645 [hep-th]].
\semi
  	C.R.~Mafra, O.~Schlotterer and S.~Stieberger,
	``Complete N-Point Superstring Disk Amplitude II. Amplitude and Hypergeometric Function Structure,''
	Nucl.\ Phys.\ B {\bf 873}, 461 (2013).
	[arXiv:1106.2646 [hep-th]].
}

\lref\wittentwistor{
	E.Witten,
        ``Twistor-Like Transform In Ten-Dimensions''
        Nucl.Phys. B {\bf 266}, 245~(1986)
}
\lref\SiegelYI{
	W.Siegel,
	``Superfields in Higher Dimensional Space-time,''
	Phys. Lett.B {\bf 80}, 220~(1979)
}
\lref\psf{
 	N.~Berkovits,
	``Super-Poincare covariant quantization of the superstring,''
	JHEP {\bf 0004}, 018 (2000)
	[arXiv:hep-th/0001035].
}
\lref\NMPS{
    N.~Berkovits,
        ``Pure spinor formalism as an N = 2 topological string,''
    JHEP {\bf 0510}, 089 (2005)
    [arXiv:hep-th/0509120].
}
\lref\MPS{
        N.~Berkovits,
	``Multiloop amplitudes and vanishing theorems using the pure spinor formalism for the superstring,''
	JHEP {\bf 0409}, 047 (2004).
	[hep-th/0406055].
}
\lref\oneloopMichael{
	M.B.~Green, C.R.~Mafra and O.~Schlotterer,
	``Multiparticle one-loop amplitudes and S-duality in closed superstring theory,''
	[arXiv:1307.3534].
}
\lref\wip{
S.~He, C.R.~Mafra and O.~Schlotterer, Work in progress
}
\lref\anomaly{
	N.~Berkovits and C.R.~Mafra,
	``Some Superstring Amplitude Computations with the Non-Minimal Pure Spinor Formalism,''
	JHEP {\bf 0611}, 079 (2006).
	[hep-th/0607187].
}
\lref\PSBCJ{
	C.R.~Mafra, O.~Schlotterer and S.~Stieberger,
	``Explicit BCJ Numerators from Pure Spinors,''
	JHEP {\bf 1107}, 092 (2011).
	[arXiv:1104.5224 [hep-th]].
}
\lref\vallette{
	J.-L. Loday and B. Vallette, ``Algebraic operads'', Grundlehren Math. Wiss. 346, Springer, Heidelberg, 2012.
}
\lref\bminusOne{J.~Polchinski,
  ``String theory. Vol. 1: An introduction to the bosonic string,''
{\it  Cambridge, UK: Univ. Pr. (1998) 402 p}
}
\lref\BCJ{
	Z.~Bern, J.J.M.~Carrasco and H.~Johansson,
	``New Relations for Gauge-Theory Amplitudes,''
	Phys.\ Rev.\ D {\bf 78}, 085011 (2008).
	[arXiv:0805.3993 [hep-ph]].
}
\lref\oldMomKer{
	Z.~Bern, L.~J.~Dixon, M.~Perelstein and J.~S.~Rozowsky,
	``Multileg one loop gravity amplitudes from gauge theory,''
	Nucl.\ Phys.\ B {\bf 546}, 423 (1999).
	[hep-th/9811140].
}
\lref\MomKer{
	N.~E.~J.~Bjerrum-Bohr, P.~H.~Damgaard, T.~Sondergaard and P.~Vanhove,
	``The Momentum Kernel of Gauge and Gravity Theories,''
	JHEP {\bf 1101}, 001 (2011).
	[arXiv:1010.3933 [hep-th]].
}
\lref\Polylogs{
	J.~Broedel, O.~Schlotterer and S.~Stieberger,
	``Polylogarithms, Multiple Zeta Values and Superstring Amplitudes,''
	Fortsch.\ Phys.\  {\bf 61}, 812 (2013).
	[arXiv:1304.7267 [hep-th]].
}
\lref\KKref{
	R.~Kleiss and H.~Kuijf,
	``Multi - Gluon Cross-sections and Five Jet Production at Hadron Colliders,''
	Nucl.\ Phys.\ B {\bf 312}, 616 (1989)..
\semi
	V.~Del Duca, L.J.~Dixon and F.~Maltoni,
	``New color decompositions for gauge amplitudes at tree and loop level,''
	Nucl.\ Phys.\ B {\bf 571}, 51 (2000).
	[hep-ph/9910563].
}
\lref\PSS{
	C.R.~Mafra,
	``PSS: A FORM Program to Evaluate Pure Spinor Superspace Expressions,''
	[arXiv:1007.4999 [hep-th]].
}
\lref\FORM{
	J.A.M.~Vermaseren,
	``New features of FORM,''
	arXiv:math-ph/0010025.
\semi
	M.~Tentyukov and J.A.M.~Vermaseren,
	``The multithreaded version of FORM,''
	arXiv:hep-ph/0702279.
}
\lref\Figueroa{
	J.M.~Figueroa-O'Farrill,
	``N=2 structures in all string theories,''
	J.\ Math.\ Phys.\  {\bf 38}, 5559 (1997).
	[hep-th/9507145].
}
\lref\GSanomaly{
	M.~B.~Green and J.~H.~Schwarz,
	``Anomaly Cancellation in Supersymmetric D=10 Gauge Theory and Superstring Theory,''
	Phys.\ Lett.\ B {\bf 149}, 117 (1984)..
\semi
	M.B.~Green and J.H.~Schwarz,
	``The Hexagon Gauge Anomaly in Type I Superstring Theory,''
	Nucl.\ Phys.\ B {\bf 255}, 93 (1985)..
}
\lref\LiEprogram{
M.A.A. van Leeuwen, A.M. Cohen and B. Lisser,
``LiE, A Package for Lie Group Computations'', Computer Algebra Nederland, Amsterdam, ISBN 90-74116-02-7, 1992
}
\lref\wipH{
C.R.~Mafra and O.~Schlotterer, work in progress.
}

\lref\MonteiroPC{
  R.~Monteiro and D.~O'Connell,
  ``The Kinematic Algebra From the Self-Dual Sector,''
JHEP {\bf 1107}, 007 (2011).
[arXiv:1105.2565 [hep-th]].
}

\lref\GomezSLA{
  H.~Gomez and C.R.~Mafra,
  ``The closed-string 3-loop amplitude and S-duality,''
[arXiv:1308.6567 [hep-th]].
}
\lref\bigHowe{
  P.S.~Howe,
  ``Pure Spinors Lines In Superspace And Ten-Dimensional Supersymmetric
  Theories,''
  Phys.\ Lett.\  B {\bf 258}, 141 (1991)
  [Addendum-ibid.\  B {\bf 259}, 511 (1991)].
\semi
  P.S.~Howe,
  ``Pure Spinors, Function Superspaces And Supergravity Theories In
  Ten-Dimensions And Eleven-Dimensions,''
  Phys.\ Lett.\  B {\bf 273}, 90 (1991).
}
\lref\NilssonCM{
  B.E.W.~Nilsson,
  ``Pure Spinors as Auxiliary Fields in the Ten-dimensional Supersymmetric {Yang-Mills} Theory,''
Class.\ Quant.\ Grav.\  {\bf 3}, L41 (1986)..
}
\lref\towards{
	C.R.~Mafra,
	``Towards Field Theory Amplitudes From the Cohomology of Pure Spinor Superspace,''
	JHEP {\bf 1011}, 096 (2010).
	[arXiv:1007.3639 [hep-th]].
}

\lref\WWW{
	C.R.~Mafra, O.~Schlotterer,
http://www.damtp.cam.ac.uk/user/crm66/SYM/pss.html
}
\lref\ambitwistor{
  L.~Mason and D.~Skinner,
  ``Ambitwistor strings and the scattering equations,''
[arXiv:1311.2564 [hep-th]].
\semi
  T.~Adamo, E.~Casali and D.~Skinner,
  ``Ambitwistor strings and the scattering equations at one loop,''
[arXiv:1312.3828 [hep-th]].
}

\lref\ChenJXA{
  Y.~-X.~Chen, Y.~-J.~Du and B.~Feng,
  ``A Proof of the Explicit Minimal-basis Expansion of Tree Amplitudes in Gauge Field Theory,''
JHEP {\bf 1102}, 112 (2011).
[arXiv:1101.0009 [hep-th]].
}

\lref\GS{
    M.~B.~Green and J.~H.~Schwarz,
    ``Covariant Description Of Superstrings,''
    Phys.\ Lett.\  B {\bf 136}, 367 (1984)
    \semi
     M.~B.~Green and J.~H.~Schwarz,
    ``Supersymmetrical String Theories,''
    Phys.\ Lett.\  B {\bf 109}, 444 (1982).
}
\lref\RNS{
    P.~Ramond,
    ``Dual Theory for Free Fermions,''
    Phys.\ Rev.\  D {\bf 3}, 2415 (1971),
    A.~Neveu and J.~H.~Schwarz,
    ``Factorizable dual model of pions,''
    Nucl.\ Phys.\  B {\bf 31} (1971) 86,
    A.~Neveu and J.~H.~Schwarz,
    ``Quark Model of Dual Pions,''
    Phys.\ Rev.\  D {\bf 4}, 1109 (1971).
}
\lref\theorems{
  N.~Berkovits,
  ``New higher-derivative R**4 theorems,''
Phys.\ Rev.\ Lett.\  {\bf 98}, 211601 (2007).
[hep-th/0609006].
}
\lref\piotr{
  A.~Ochirov and P.~Tourkine,
  ``BCJ duality and double copy in the closed string sector,''
[arXiv:1312.1326 [hep-th]].
}
\lref\sixtree{
  C.~R.~Mafra, O.~Schlotterer, S.~Stieberger and D.~Tsimpis,
  ``Six Open String Disk Amplitude in Pure Spinor Superspace,''
Nucl.\ Phys.\ B {\bf 846}, 359 (2011).
[arXiv:1011.0994 [hep-th]].
}
\lref\dennen{
  Z.~Bern and T.~Dennen,
  ``A Color Dual Form for Gauge-Theory Amplitudes,''
Phys.\ Rev.\ Lett.\  {\bf 107}, 081601 (2011).
[arXiv:1103.0312 [hep-th]].
}
\lref\moore{
  G.~W.~Moore,
  ``Symmetries of the bosonic string S matrix,''
[hep-th/9310026].
}
\lref\lian{
  B.~H.~Lian and G.~J.~Zuckerman,
  ``New perspectives on the BRST algebraic structure of string theory,''
Commun.\ Math.\ Phys.\  {\bf 154}, 613 (1993).
[hep-th/9211072].
}
\lref\GreenFT{
  M.~B.~Green, J.~H.~Schwarz and L.~Brink,
  ``N=4 Yang-Mills and N=8 Supergravity as Limits of String Theories,''
Nucl.\ Phys.\ B {\bf 198}, 474 (1982).
}
\lref\siegel{
	W.~Siegel,
	``Classical Superstring Mechanics,''
	Nucl.\ Phys.\  {\bf B263}, 93 (1986).
}
\lref\BCJloop{
  Z.~Bern, J.~J.~M.~Carrasco and H.~Johansson,
  ``Perturbative Quantum Gravity as a Double Copy of Gauge Theory,''
Phys.\ Rev.\ Lett.\  {\bf 105}, 061602 (2010).
[arXiv:1004.0476 [hep-th]].
}

\lref\MonteiroRYA{
  R.~Monteiro and D.~O'Connell,
  ``The Kinematic Algebras from the Scattering Equations,''
JHEP {\bf 1403}, 110 (2014).
[arXiv:1311.1151 [hep-th]].
}

\listtoc
\writetoc
\filbreak

\newsec Introduction

Many examples have shown that String Theory inspires a deeper understanding of scattering amplitudes in field theories, see e.g.
\refs{\GreenFT,\KawaiXQ, \BjerrumBohrRD, \StiebergerHQ}. The world--sheet viewpoint on point-particle interactions offers useful guiding
principles through the multitude of Feynman diagrams. For example, tree-level subdiagrams of external particles arise
when insertion points of string states on the world--sheet collide. This is captured by the operator product expansion
(OPE) among vertex operators.

In this work, we study this mechanism in the context of ten-dimensional super Yang--Mills (SYM) theory. Its superspace
description benefits from the use of pure spinors \refs{\NilssonCM,\bigHowe}, and this formulation directly descends from the pure spinor
superstring \psf. In previous work, a family of so-called BRST building blocks was identified in the pure spinor
formalism \refs{\nptMethod, \nptTree} which encompasses the superfield degrees of freedom of several external
particles. These BRST blocks were argued to represent tree-level subdiagrams and led to an elegant and manifestly supersymmetric solution for multileg tree-level amplitudes in SYM theory \nptMethod\ and the full-fledged open superstring\foot{See \BarreiroDPA\ for an indirect derivation of open superstring trees among gluons, based on bosonic gauge invariance and kinematic constraints from the RNS worldsheet prescription.} \refs{\nptTree}.
As initially suggested in \towards, the driving forces in these constructions were:
\medskip
\item{(i)} The (iterated) OPE of gluon multiplet vertex operators

\item{(ii)} The action of the BRST operator on the OPE output to identify the symmetry components in the cohomology

\item{(iii)} BRST-invariance of the full amplitude dictates the composition of BRST-covariant tree diagrams

\noindent In step (ii), we benefit from the simple form of the BRST action on kinematic degrees of freedom, based on the SYM equations of motion
for the superfields \refs{\wittentwistor,\SiegelYI}. This appears to be special to the pure spinor formalism, at least we are not
aware of an analogous implementation in the Ramond-Neveu-Schwarz (RNS) \RNS\ or Green-Schwarz (GS) \GS\ framework.

The tree-level setup of \refs{\nptMethod, \nptTree} only made use of the mixed OPEs between one unintegrated vertex operator $V$ and one integrated
version $U$. In recent one- and three-loop-calculations \refs{\oneloopbb, \oneloopMichael, \GomezSLA}, on the other
hand, it became clear that pieces of the OPE among $U$ vertices had similar covariant BRST properties leading to a
simplified description of the amplitudes. In the
following, we will complete the list of such BRST-covariant OPE ingredients and introduce multiparticle versions of
the integrated vertex operator.

The multiparticle vertex operators are defined in terms of multiparticle superfields of ten-dimensional
SYM theory. The latter in turn are constructed recursively where the rule for adding particles is extracted
from the OPE among single particle vertex operators.
The BRST transformation of
these vertex operators is equivalent to equations of motion for the multiparticle superfields, which take the
same form as their single-particle counterparts \refs{\wittentwistor,\SiegelYI}, but are enriched by contact terms. It
points to very fundamental structures of SYM theory that these combinations of single-particle
fields reproduce the ``elementary'' equations of motions.

In more mathematical terms \refs{\vallette, \mathA, \loday}, the recursion rule fusing two multiparticle superfields
to a larger representative can be viewed as a Lie bracket
operation which implements the algebraic structure of tree level graphs. In particular, the aforementioned
contact terms present in multiparticle equations of motion directly realize the Lie symmetries of tree subdiagrams.
This carries the flavour of a kinematic algebra which
might shed further light on the duality between color and kinematics \BCJ\ in ten dimensions\foot{See
\refs{\MonteiroPC,\MonteiroRYA} for related work on the kinematic algebra in four and arbitrary dimensions.}.
More specifically, the Lie symmetries of multiparticle BRST blocks imply kinematic Jacobi relations within the
corresponding tree subdiagrams.

The multiparticle superfields and their BRST properties turn out to guide the construction of BRST-invariant kinematic
factors. Together with the tight contraints from zero-mode saturation, this allows to anticipate the structure of
scattering amplitudes in both field theory and string theory. As an example, we conclude this paper with an application
to one-loop amplitudes of the open and closed (type II) superstring. The
pure spinor formulation of the five graviton amplitude in \oneloopMichael\ gave an example of how vector contractions
between left- and right-moving superfields can be implemented in a BRST-invariant way. The backbone of this superspace
construction is a vectorial BRST cohomology element which we recursively extend to higher multiplicity. 
From the field theory perspective, this
amounts to identifying loop momentum dependent parts of the numerators, see \refs{\piotr,\wip}.

The limit of infinite string tension $\alpha' \rightarrow 0$ leads to a worldline realization of the pure spinor setup
\WLA\ (see also \ambitwistor\ for the RNS equivalent). It has been shown in \WLB\ that the worldline modifications of the worldsheet vertex operators and their OPEs
give rise to the same SYM tree amplitudes as previously obtained from superstring methods \refs{\nptMethod,\nptTree}.
Accordingly, it would be interesting to find the worldline equivalent of the present BRST block constructions.

\newsec Review

\subsec Ten-dimensional SYM theory

Linearized super-Yang--Mills theory in ten dimensions can be described using the
superfields\foot{It is customary to use a calligraphic letter for the superfield field-strength. However in this
paper calligraphic letters will denote the Berends--Giele currents associated to the superfields, see section 4.}
$A_\a(x,\t)$, $A_m(x,\t)$, $W^\a(x,\t)$ and $F_{mn}(x,\t)$ satisfying \refs{\wittentwistor,\SiegelYI}
\eqn\RankOneEOM{
\eqalign{
2 D_{(\a} A_{\b)} & = \g^m_{\a\b} A_m\cr
D_\a F_{mn} & = 2k_{[m} (\g_{n]} W)_\a
}\qquad\eqalign{
D_\a A_m &= (\g_m W)_\a + k_m A_\a  \cr
D_\a W^{\b} &= {1\over 4}(\g^{mn})^{\phantom{m}\b}_\a F_{mn}.
}}
with gauge transformations $\d A_\a = D_\a \Omega$ and $\d A_m = k_m \Omega$ for any superfield $\Omega$.
The above equations of motion imply that the superfields $A_m$, $W^\a$ and $F^{mn}$ can be derived
from the spinor superpotential $A_\a$,
\eqn\Derived{
A^m = {1\over 8}(D\g^m A),\quad W^\a = -{1\over 10}\g^{\a\b}_m(k^m A_\b - D_\b A^m),\quad
F_{mn} = {1\over 8}(\g_{mn})_\a{}^\b D_\b W^\a\,.
}
The notion that the superfield $A_\a$ is enough to derive the others will be used in the next
section to obtain a multiparticle generalization of the above equations of motion.

\subsec BRST building blocks from vertex operators

In the pure spinor formalism the massless sector of the open superstring (i.e. the gluon multiplet) is described by the vertex
operators
\eqn\Uintegrated{
V= \l^\a A_\a, \qquad U = \p\t^\a A_\a + \Pi^m A_m + d_\a W^\a + \half N^{mn}F_{mn} \ .
}
The superfields $K \in \{A_\a, A_m , W^\a ,F_{mn} \}$ and the pure spinor ghost $ \l^\a$ carry conformal
weight zero whereas the worldsheet fields $\{\p\t^\a , \Pi^m , d_\a ,N^{mn}\}$ have conformal weight one.
When the superfields are on-shell and the pure spinor constraint $(\l \gamma^m \l)=0$ is
imposed, the vertices satisfy \psf
\eqn\PSEOM{
QV=0,\qquad QU = \p V\,,
}
where $Q=\l^\a D_\a$ is the BRST charge. The above fields obey the following OPEs \refs{\siegel,\psf},
\eqn\allOPEs{
\eqalign{
d_\a(z_i) K(z_j)      &\rightarrow  {D_\a K\over z_{ij}},\cr
d_\a(z_i) \Pi^m(z_j) &\rightarrow  {(\g^m\p\t)_\a\over z_{ij}}\cr
d_\a(z_i)\t^\b(z_j)  &\rightarrow  {\d^\b_\a\over z_{ij}}\cr
d_\a(z_i)\p\t^\b(z_j) &\rightarrow  {\d^\b_\a\over z_{ij}^2},\cr
}\quad\eqalign{
\Pi^m(z_i) K(z_j)    &\rightarrow -  {k^m K\over z_{ij}},\cr
\Pi^m(z_i)\Pi^n(z_j) &\rightarrow -  {\eta^{mn}\over z_{ij}^2}, \cr
d_\a(z_i)d_\b(z_j)    &\rightarrow -  {\g^m_{\a\b}\Pi_m\over z_{ij}},\cr
N^{mn}(z_i)\l^\a(z_j) &\rightarrow - \half {(\l\g^{mn})^\a\over z_{ij}}\cr
}}
and
\eqn\NNOPE{
N^{mn}(z_i)N_{pq}(z_j) \rightarrow  {4 \over z_{ij}}N^{[m}_{\phantom{m}[p}\d^{n]}_{q]}
- {6 \over z_{ij}^2}\d^n_{[p}\d^m_{q]}\,,
}
where $z_{ij}=z_i-z_j$ are worldsheet positions. By $K(x,\t)$, we collectively denote any superfield containing only zero-modes of $\t^\a$ and whose $x$ dependence
is entirely given by the plane wave factor\foot{To avoid factors of $i$ in the formulae,
we define $ik^m\equiv k^m$.} $e^{k\cdot x}$.

Starting with the recursive definition of
\eqn\Ldefs{
\lim_{z_2\to z_1} V^1(z_1)U^2(z_2) \rightarrow {L_{21}\over z_{21}}, \quad
\lim_{z_p \to z_1} L_{2131...(p-1)1}(z_1) U^p(z_p) \rightarrow {L_{2131...(p-1)1p1}\over z_{p1} },
}
fermionic ghost-number one {\it BRST building blocks} were defined in \refs{\nptMethod,\nptTree} by removal of BRST exact
terms,
\eqn\Tdef{
T_{123\ldots p} = L_{2131\ldots p} - Q( \ldots).
}
They transform covariantly under BRST variation, for instance
$$
QT_{12} = (k^1\cdot k^2) T_1 T_2,\qquad QT_{123} = (k^1\cdot k^2)(T_1 T_{23} + T_{13}T_2) + (k^{12}\cdot k^3)T_{12}T_3
$$
at rank two and three. More generally,
\eqn\BRSTvar{
QT_{12{\dots} p} = \sum_{j=2}^{p} \sum_{\a \in P(\beta_j)} (k^{12\ldots j-1}\cdot k^j)\,
T_{12\ldots  j-1,\{\a\}}\; T_{j, \{\beta_j \backslash \a\}},
}
where $\b_j = \{j+1,\dots, p\}$ and $P(\b_j)$ is the powerset of $\b_j$. Moreover, we identify $T_i \equiv V_i$
for a single-particle label $i$
and abbreviate multiparticle momenta by $k^{123\ldots p}_m \equiv \sum_{i=1}^p k^i_m$.

\ifig\figone{The correspondence of cubic graphs and BRST building blocks.}
{\epsfxsize=0.60\hsize\epsfbox{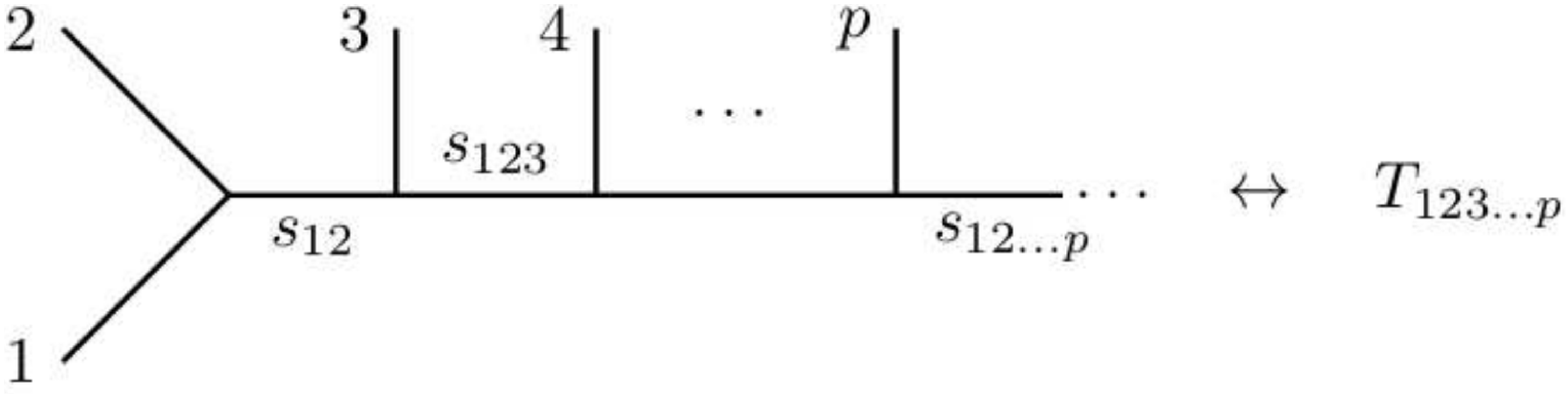}}

\subsec Lie symmetries of BRST building blocks

After removal of $Q$ exact terms in \Tdef, BRST building blocks $T_{12 \ldots p}$ satisfy all the Lie symmetries ${\lie}_k$ of
tree-level graphs for $2\le k \le p$, where\foot{Throughout this work, antisymmetrization over $N$ labels associated with external particles (as in \BRSTsymold) does not contain
an overall $1/N!$. However, antisymmetrized Lorentz indices $m,n,p,\ldots$ are presented in the convention $A_{[mn]}={1\over 2}(A_{mn}-A_{nm})$.}
\eqnn\BRSTsymold
$$\eqalignno{
\lie_{k=2n+1}\colon\quad&{} T_{12\ldots n+1[n+2[\ldots [2n-1[2n,2n+1]] \ldots ]]} - T_{2n+1\ldots n+2[n+1[\ldots [3[21]] \ldots ]]}  = 0 \cr
\lie_{k=2n}\colon\quad&{} T_{12\ldots n[n+1[\ldots [2n-2[2n-1,2n]] \ldots ]]} + T_{2n\ldots n+1[n[\ldots [3[21]] \ldots ]]}  = 0.&\BRSTsymold\cr
}$$
Defining the operator $\lie_k\circ$ as the ``Lie symmetry generator'', the first few examples of the symmetries
\BRSTsymold\ are
\eqnn\LieEx
$$\eqalignno{
0 &= \lie_2\circ T_{12} \equiv T_{12}+T_{21}, \cr
0 &= \lie_3\circ T_{123} \equiv T_{123} + T_{231} + T_{312},&\LieEx\cr
0 &= \lie_4\circ T_{1234} \equiv T_{1234} - T_{1243} + T_{3412} - T_{3421}.\cr
}$$
The symmetries \BRSTsymold\ have been denoted ``Lie'' because a contraction
of Lie algebra structure constants satisfies the same symmetries \oneloopbb\foot{Under the {\it Dynkin bracketing} operation, the building blocks satisfy
$T_{[[\ldots[[1,2],3],\ldots],p]} = p T_{123 \ldots p}$ and therefore
they belong to ${\rm Lie}(p)$. See e.g.\ Proposition 13.2.3 of \vallette.},
\eqn\struCte{
T_{1234 \ldots p} \leftrightarrow\; f^{12 a_2} \, f^{a_2 3 a_3} \, f^{a_3 4 a_4} \ldots f^{a_{p-1} p a_p}\,
}
and therefore the building blocks have the correct behavior to describe the kinematic numerators
of cubic graphs, see \figone.

\subsec Lie symmetries versus BRST variations

It is crucial to notice the interplay between the BRST variations \BRSTvar\ and the Lie symmetries \BRSTsymold\ of
cubic tree level subdiagrams: At rank two and three, we have
\eqn\NumberTs{
Q(T_{12} + T_{21}) = 0,\qquad Q(T_{123} + T_{213}) = Q(T_{123} + T_{231} + T_{312}) = 0 \ ,
}
and the BRST variation \BRSTvar\ always has the precise form to make the sums in \BRSTsymold\ BRST closed. This closure
even holds before the redefinitions \Tdef\ are performed, e.g. $Q (L_{12} + L_{21}) = 0$ for the direct outcome of the
OPE \Ldefs. Any such BRST closed combination is also BRST exact since its conformal weight $\sim k_{12\ldots p}^2$ is different from zero (unless $p=1$)\foot{Recall that in a
topological conformal field theory $Qb_0 = L_0$ implies that if $Q\phi_h =0$ and $L_0 \phi_h = h \phi_h$, then $\phi_h
= (1/h)Q(b_0\phi_h)$ for $h\neq 0$. See e.g.\ \Figueroa.}. As detailed in \refs{\nptMethod,\nptTree}, this implies
that BRST exact terms (such as $Q(A_1\cdot A_2)= L_{21}+L_{12})$) can be subtracted in the definition of
$T_{12\ldots p}$ given in \Tdef. Therefore the Lie symmetries obeyed by $T_{12 \ldots p}$ are a
consequence of the underlying BRST cohomology nature of the pure spinor superspace expressions which will
ultimately describe the scattering amplitudes.

However, it was a matter of trial and error to find the BRST-``ancestors'' of
$Q$-closed $L_{21\ldots p1}$ combinations, such as $(A_1 \cdot A_2)$ in the rank-two example and more lengthy expression
at rank $\leq 5$ given in \nptTree. In the following section, we develop a constructive method to generate these BRST
completions in \Tdef\ without any guesswork. Moreover, our current approach based on integrated vertex operators $U_i$
delays the need for redefinitions \Tdef\ to rank three; all the rank-two BRST blocks will automatically be antisymmetric
since they follow from the simple pole of the OPE between two integrated vertices.


The BRST building blocks play a key role in the recursive BRST cohomology method to compute SYM tree-level amplitudes
\refs{\nptMethod,\towards} and in obtaining a manifestly local representation of BCJ-satisfying \BCJ\ tree-level numerators \PSBCJ.
However, their explicit superspace expressions in \nptTree\ following from the more and more cumbersome OPE computations
\Ldefs\ become lengthy for higher rank and lack an organizing principle. We will describe a recursive method in the next section to
find compact representations and to completely bypass the CFT calculations beyond rank two.

\ifig\figtwo{Four superfield realizations $K_B\in \{A^B_\a, A^B_m , W_B^\a ,F^B_{mn}\}$ of cubic
tree graphs $B = b_1b_2\ldots b_p$. This generalizes the mapping in fig.~1 from previous
work \refs{\nptMethod,\nptTree} where only one representative $T_B$ at ghost number one was given.}
{\epsfxsize=0.60\hsize\epsfbox{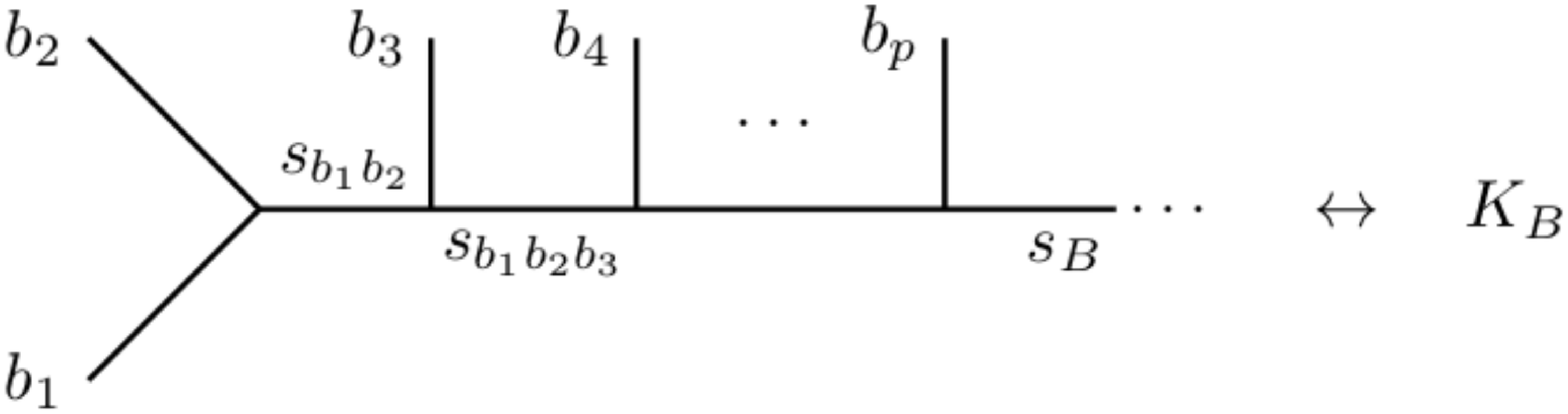}}

\newsec Pure spinor BRST blocks

\seclab\BRSTsec

\noindent In this section we will show how to recursively define multiparticle superfields $A_\a^B(x,\t)$, $A^m_B(x,\t)$,
$W^\a_B(x,\t)$ and $F^{mn}_B(x,\t)$. As we will see, the recursion is driven by the OPE among two single-particle vertex operators and a system of multiparticle SYM equations of motion
which generalize the standard description of \RankOneEOM.
Throughout this paper, upper case letters from the beginning of the Latin alphabet
will represent multiparticle labels, e.g. $B=b_1b_2{\ldots} b_p$ at rank $p \equiv |B|$. In particular, whenever they are attached to a
multiparticle superfield $K_B \in \{ A^B_\a, A^B_m , W_B^\a ,F^B_{mn} \}$ (without any hats or primes), the $B=b_1b_2
{\ldots} b_p$ carry the same Lie symmetries \BRSTsymold,
\eqnn\BRSTsym
$$\eqalignno{
\lie_{k=2n+1}\colon\quad&{} K_{12\ldots n+1[n+2[\ldots [2n-1[2n,2n+1]] \ldots ]]} - K_{2n+1\ldots n+2[n+1[\ldots [3[21]] \ldots ]]}  = 0 \cr
\lie_{k=2n}\colon\quad&{} K_{12\ldots n[n+1[\ldots [2n-2[2n-1,2n]] \ldots ]]} + K_{2n\ldots n+1[n[\ldots [3[21]] \ldots ]]}  = 0.&\BRSTsym\cr
}$$
The superfields $\{ A^B_\a, A^B_m , W_B^\a ,F^B_{mn} \}$ of multiplicity $p\equiv |B|$ satisfying all the Lie symmetries
$\lie_k$ for $k\le p$ will be referred to as {\it BRST blocks}\foot{Throughout this paper, we will distinguish {\it BRST building blocks} $T_B$ as
reviewed in section~2 from {\it BRST blocks} $K_B \in \{ A^B_\a, A^B_m , W_B^\a ,F^B_{mn} \}$ to be constructed in this section.}. Given
the symmetry matching relation
\eqn\newstruCte{
K_{1234 \ldots p} \leftrightarrow\; f^{12 a_2} \, f^{a_2 3 a_3} \, f^{a_3 4 a_4} \ldots f^{a_{p-1} p a_p}\,
}
with color factors,
the BRST blocks reproduce symmetry properties of Lie algebraic structure constants. The BCJ compatibility of the explicit
tree-level numerators in \PSBCJ\ are based on $\lambda^\a A^B_\a$ satisfying this symmetry matching. As described in
the mathematics literature \refs{\vallette, \mathA}, the associated cubic graphs shown in \figtwo\ (planar binary
trees in mathematical jargon) can be mapped to iterated brackets and thereby give rise to a general construction of a
Lie algebra basis. More details are given in Appendix~A.



The BRST variation of the multiparticle unintegrated vertex operator defined by $V^B \equiv \l^\a A^B_\a$ will be shown to
have the same functional form as the BRST variation \BRSTvar\ of $T_B$, thereby constituting a new representation of such objects. BRST-invariants built from $T_B$ do not change under a global redefinition $T_B \rightarrow V_B$, hence the representations are equivalent. From now on, $T_B$ from \refs{\nptMethod,\nptTree} will not be used anymore and the new representation $V_B$ will take its place because it follows from simpler principles. 

\subsec Rank two

The way towards multiparticle BRST blocks is suggested by the OPE between two integrated vertex operators. This is the largest and only CFT computation relevant for this work and has
been firstly performed in \fivetree,
\eqnn\UUope
$$\eqalignno{
U^1(z_1) U^2(z_2) \rightarrow & z_{12}^{-k^1 \cdot k^2-1} \Big( \partial \theta^{\alpha} \big[ (k^1 \cdot A_2) A^1_{\alpha} -(k^2 \cdot A_1) A^2_{\alpha} + D_{\alpha} A_{\beta}^2 W_1^{\beta} - D_{\alpha} A_{\beta}^1 W_2^{\beta} \big] \cr
+& \Pi^m \big[ (k^1 \cdot A_2) A_m^1 - (k^2 \cdot A_1) A_m^2 + k_m^2 (A_2 W_1) -  k_m^1 (A_1 W_2) - (W_1 \gamma_m W_2)\big]\cr
+ & d_{\alpha} \big[ (k^1 \cdot A_2) W_1^{\alpha} - (k^2 \cdot A_1) W_2^{\alpha} + {1\over 4} (\gamma^{mn} W_1)^{\alpha} F_{mn}^2- {1\over 4} (\gamma^{mn} W_2)^{\alpha} F_{mn}^1 \big]\cr
+ & {1 \over 2} N^{mn} \big[ (k^1 \cdot A_2) F^1_{mn} - (k^2 \cdot A_1) F^2_{mn} - 2 k_{m}^{12} (W_1 \gamma_n W_2) - 2 F_{ma}^2 F^{3 \ a}_n \big] \Big)\cr
+ & (1+k^1 \cdot k^2) z_{12}^{-k^1 \cdot k^2-2} \big[(A_1 W_2) + (A_2 W_1) - (A_1 \cdot A_2) \big].&\UUope
}$$
Using the relation $\partial K = \partial \theta^{\alpha} D_{\alpha} K + \Pi^m k_m K$ for superfields $K$ independent on $\partial \theta^{\alpha}$ and $\l^\a$,
we can absorb the most singular piece $\sim z_{12}^{-k^1 \cdot k^2-2}$ into total $z_{1},z_2$ derivatives and rewrite
\eqnn\UUUope
$$\eqalignno{
U^1(z_1) U^2(z_2) \rightarrow&
- z_{12}^{-k^1 \cdot k^2-1} \big[ \p\t^\a A^{12}_\a + \Pi^m A^{12}_m + d_\a W^\a_{12} + \half N^{mn}F^{12}_{mn} \big] &\UUUope\cr
+\partial_1 \Big( z_{12}^{-k^1 \cdot k^2-1} &\big[ {1\over 2} (A_1 \cdot A_2) - (A_1 W_2) \big] \Big)
- \partial_2 \Big( z_{12}^{-k^1 \cdot k^2-1} \big[ {1\over 2} (A_1 \cdot A_2) - (A_2 W_1) \big] \Big)
}$$
where
\eqnn\Atwo
$$\eqalignno{
A^{12}_\a &= - \half\bigl[ A^1_\a (k^1\cdot A^2) + A^1_m (\g^m W^2)_\a - (1\leftrightarrow 2)\bigr]
&\Atwo \cr
A^{12}_m &=  \half\Bigl[ A^1_p F^2_{pm} - A^1_m(k^1\cdot A^2) + (W^1\g_m W^2) - (1\leftrightarrow 2)\Bigr] \cr
W_{12}^\a &= {1\over 4}(\g^{mn}W^2)^\a F^1_{mn} + W_2^\a (k^2\cdot A^1) - (1\leftrightarrow 2)\cr
F^{12}_{mn} &= F^2_{mn}(k^2\cdot A^1) +  F^2_{[m}{}^{p}F_{n]p}^1 + k_{12}^{[m}(W_1\g^{n]}W_2) - (1\leftrightarrow 2)\cr
& = k^{12}_m A^{12}_n - k^{12}_n A^{12}_m - (k^1\cdot k^2)(A^1_m A^2_n -A^1_n A^2_m). \cr
}$$
Note that the last line can be viewed as a multiparticle generalization of the field-strength relation $F^i_{mn}=k^i_m A_n - k^i_n A_m$,
modified by the contact terms $(k^1\cdot k^2)(A^1_m A^2_n -A^1_n A^2_m)$.

In the prescription for computing string amplitudes the vertex operators are integrated over the worldsheet so the
total derivatives can be dropped\foot{In string calculations this cancellation involves a subtle interplay of
BRST-exact terms and total derivatives on the worldsheet, see \towards\ and \sixtree\ for  five- and six-point examples at tree level.
One manifestation is the agreement of the superfields along with $\partial_1, \partial_2$ in \UUUope\ with the BRST-exact admixtures in
$V_1(z_1) U_2(z_2) \rightarrow z_{12}^{-k^1\cdot k^2-1}(V_{12}+ Q[(A_1 W_2) - {1\over 2}(A_1 \cdot A_2)] )$.
} and the composite superfields in \Atwo\ can be picked up via
\eqnn\commutator
$$\eqalignno{
U^{12} & =  - \oint z_{12}^{k^1 \cdot k^2} U^1(z_1) U^2(z_2) &\commutator\cr
&= 
\p\t^\a A^{12}_\a + \Pi^m A^{12}_m + d_\a W^\a_{12} + \half N^{mn}F^{12}_{mn}.
}$$
One can check using \RankOneEOM\ that the above superfields satisfy
\eqnn\EOMAtwo
\eqnn\EOMAmtwo
\eqnn\EOMWtwo
\eqnn\EOMFtwo
$$\eqalignno{
2 D_{(\a} A^{12}_{\b)} &= \g^m_{\a\b}A^{12}_m + (k^1\cdot k^2)(A^1_\a A^2_\b + A^1_\b A^2_\a) &\EOMAtwo\cr
D_\a A^{12}_m &= (\g_m W^{12})_\a + k^{12}_m A^{12}_\a + (k^1\cdot k^2)(A^1_\a A^2_m - A^2_\a A^1_m) &\EOMAmtwo\cr
D_\a W^\b_{12} &= {1\over 4}(\g^{mn})_\a{}^\b F^{12}_{mn} + (k^1\cdot k^2)(A^1_\a W_2^\b - A^2_\a W^\b_1) &\EOMWtwo\cr
D_\a F^{12}_{mn} &= k^{12}_m (\g_n W^{12})_\a - k^{12}_n (\g_m W^{12})_\a + (k^1\cdot k^2)(A^1_\a F^2_{mn} - A^2_\a
F^1_{mn}) &\EOMFtwo \cr
& \ \  + (k^1\cdot k^2)( A^{1}_{n} (\g_{m} W^2)_\a - A^{2}_{n} (\g_{m} W^1)_\a - A^{1}_{m} (\g_{n} W^2)_\a + A^{2}_{m} (\g_{n} W^1)_\a),\cr
}$$
which is a clear generalization of the standard equations of motion \RankOneEOM\ with corrections
proportional to the conformal weight $\sim {1 \over 2}(k^1+k^2)^2 = (k^1\cdot k^2)$ of the superfields. Furthermore,
the single-particle relations $k^m A^i_m=0$ and $k_m (\g^m W_i)_\a = 0$ imply that,
\eqnn\Amtwoonshell
\eqnn\Wtwoonshell
\eqnn\Ftwoonshell
$$\eqalignno{
k_{12}^m A^{12}_m &= 0 &\Amtwoonshell\cr
k^{12}_m (\g^m W^{12})_\a &= (k^1\cdot k^2)\bigl[ A^1_m (\g^m W^2)_\a - (1\leftrightarrow 2)\bigr] &\Wtwoonshell\cr
k_{12}^m F^{12}_{mn} &= (k^1\cdot k^2)\bigl[A^{12}_n + A^1_n (k^1\cdot A^2) - (1\leftrightarrow 2)\bigr] .&\Ftwoonshell\cr
}$$
In other words, the (supersymmetrized) Dirac and YM equations $k^{i}_m (\g^m W^{i})_\a=0$ and $k_{i}^m F^{i}_{mn} =0$ for single-particle superfields are modified by the same kind of contact term $\sim (k^1\cdot k^2)$ as the field strength relation in \Atwo\ and the equations of motion \EOMAtwo\ to \EOMFtwo.

Defining the rank-two unintegrated vertex operator as
\eqn\Vtwodef{
V^{12} = \l^\a A^{12}_\a
}
analogously to $V^{i} = \l^\a A^{i}_\a$, one can show that
\eqnn\QVtwo
\eqnn\QUtwo
$$\eqalignno{
Q V^{12} &= (k^1\cdot k^2) V_1 V_2 &\QVtwo\cr
QU^{12} &= \p V^{12} + (k^1\cdot k^2)(V^1 U^2 - V^2 U^1)\,, &\QUtwo
}$$
which generalizes \PSEOM\ by contact terms and reproduces the BRST variation of the
building block $T_{12}$ of \nptMethod. It is interesting to note that \QUtwo\ is compatible
with the standard prescription relating integrated and unintegrated vertices, $U^{12} = b_{-1}V^{12}$ \bminusOne.

Note that all rank-two BRST blocks are antisymmetric and therefore $U^{12} = -U^{21}$.

\subsec Rank three

Since the rank-two BRST blocks obey generalized SYM equations of motion one is tempted
to define the rank-three BRST blocks following a similar approach. We know from \RankOneEOM\ that the
standard superfields $A_m$, $W^\a$ and $F^{mn}$ can be obtained from the spinor superpotential
$A_\a$ by recursively computing covariant derivatives. We will show that the a similar approach can
be used to obtain their multiparticle generalizations starting from the following ansatz
for the superpotential,
\eqn\Athree{
\hat A^{123}_\a = - \half\bigl[ A^{12}_\a (k^{12}\cdot A^3) + A^{12}_m (\g^m W^3)_\a - (12\leftrightarrow 3)\bigr]\,.
}
This is a direct generalization of the expression for $A_{\a}^{12}$ in \Atwo\ as obtained from the OPE of
$U^{1}(z_1)U^2(z_2)$. We have now inserted two-particle data represented by $A^{12}_\a,k^{12}_m,A^{12}_m$ and
$W^\a_{12}$ into the OPE-inspired recursion. Once the
BRST-trivial symmetry components are subtracted from $\hat A^{123}_{\a}$ (see section 3.2.1), the definition \Athree\
can be interpreted in terms of a ``grafting'' procedure defined for example in \loday. As illustrated in \figgraft,
\Athree\ amounts to adjoining a further leg to the cubic graph associated with the BRST blocks $K_{12}$ at rank two,
see Appendix A for more details.

\ifig\figgraft{The essentials of the first rank three BRST block $K_{123} \in \{ A_\a^{123}, \ldots \}$ are captured
by combining $K_{12} \in \{ A_\a^{12}, A^m_{12}, W^\a_{12}\}$ and $K_{3} \in \{ A_\a^{3}, A^m_{3}, W^\a_{3} \}$. At
the level of diagrams, this can be interpreted as grafting the trees associated with $K_{12}$ and $K_3$.}
{\epsfxsize=0.60\hsize\epsfbox{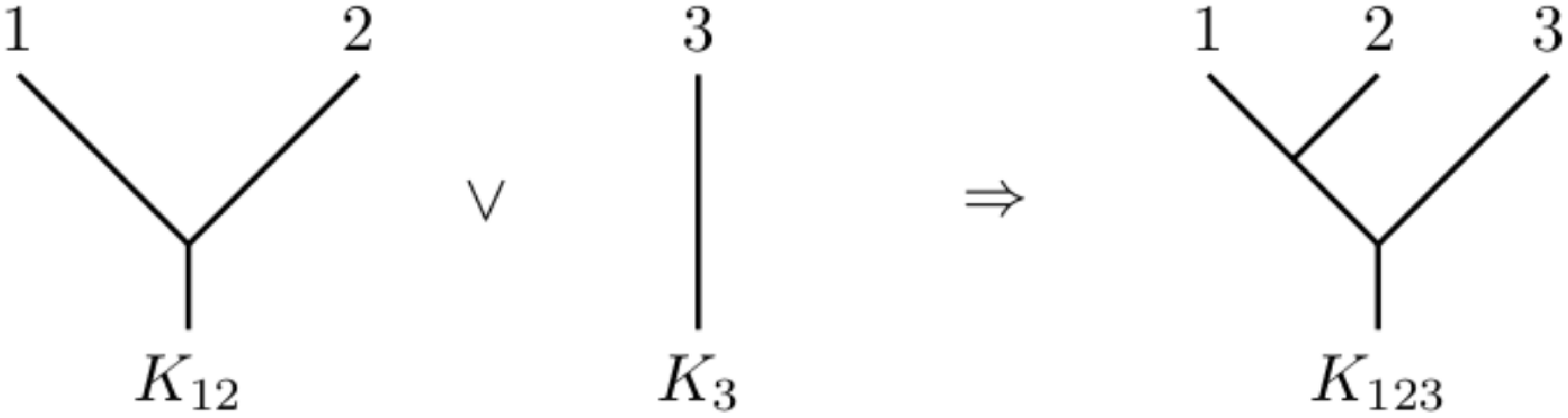}}


A short computation shows that the action of the covariant derivative can be written in a form
similar to \EOMAmtwo\ and therefore can be used to define $\hat A_{123}^m$,
\eqnn\EOMAthree
$$\eqalignno{
D_{\a}\hat A^{123}_{\b}  + D_{\b}\hat A^{123}_{\a} &= \g^m_{\a\b}\hat A^{123}_m \cr
&\quad{} + (k^1\cdot k^2)\bigl[A^1_\a A^{23}_\b + A^{13}_\a A^2_\b - (1\leftrightarrow 2)\bigr]\cr
&\quad{} + (k^{12}\cdot k^3)\bigl[A^{12}_\a A^3_\b - (12\leftrightarrow 3)\bigr]  &\EOMAthree\cr
}$$
where
\eqn\Amthree{
\hat A^{123}_m = \half\Bigl[ A_{12}^p F^3_{pm} - A^{12}_m (k^{12}\cdot A^3) + (W^{12}\g_m W^3) - (12\leftrightarrow 3)\Bigr]\,.
}
In turn, computing the covariant derivative of \Amthree\ and rewriting the result
in a form analogous to the standard equation of motion for $A^m$ leads to the
definition of $W_{123}^\a$,
\eqnn\EOMAmthree
$$\eqalignno{
D_\a \hat A^{123}_m &= (\g_m W^{123})_\a + k^{123}_m \hat A^{123}_\a\cr
&\quad{} + (k^1\cdot k^2)\bigl[ A^1_\a A^{23}_m + A^{13}_\a A^2_m - A^{23}_\a A^1_m - A^2_\a A^{13}_m\bigr] &\EOMAmthree\cr
&\quad {}+ (k^{12}\cdot k^3)(A^{12}_\a A^3_m - A^3_\a A^{12}_m)\cr
}$$
where
\eqnn\Wthree
$$\eqalignno{
W_{123}^\a &= - (k^{12}\cdot A^3) W_{12}^\a +{1\over 4}(\g^{rs}W^3)^\a F^{12}_{rs} - (12 \leftrightarrow 3) \cr
&\quad {}+ \half (k^1\cdot k^2) \bigl[ W_2^\a (A^1\cdot A^3) - (1 \leftrightarrow 2)\bigr]\,.&\Wthree\cr
}$$
Computing the covariant derivative of \Wthree\ leads to the definition of $F^{123}_{mn}$,
\eqnn\EOMWthree
$$\eqalignno{
D_\a W^\b_{123} &= {1\over 4}(\g^{mn})_\a{}^\b F^{123}_{mn} &\EOMWthree\cr
&\quad{} + (k^1\cdot k^2)\big[A^1_\a W_{23}^\b + A^{13}_\a W_2^\b - (1 \leftrightarrow 2)\bigr]\cr
&\quad{} + (k^{12}\cdot k^3) \bigl[ A^{12}_\a W_3^\b - (12 \leftrightarrow 3)\bigr]\,,
}$$
where \Wtwoonshell\ has been used to arrive at,
\eqnn\Fthree
$$\eqalignno{
F^{123}_{mn} &= (k^3\cdot A^{12})F^3_{mn} + F^{12}_{a[m}F^3_{n]a} + 2k^{12}_{[m}(W^3 \g_{n]}W^{12}) - (12 \leftrightarrow 3)\cr
&\quad{} + (k^1\cdot k^2)\Bigl[ \half F^2_{mn}(A^1\cdot A^3) + 2A^1_{[m}(W^3\g_{n]}W^2) - (1\leftrightarrow 2)\Bigr]. & \Fthree\cr
}$$
And finally,
\eqnn\DFthree
$$\eqalignno{
D_\a F^{123}_{mn} &= 2 k^{123}_{[m} (\g_{n]} W^{123})_\a \cr
& + (k^{1}\cdot k^2)\bigl[ A^1_\a F^{23}_{mn} + A^{13}_\a F^2_{mn} - (1\leftrightarrow 2)\big]\cr
& + (k^{12}\cdot k^3)\bigl[ A^{12}_\a F^3_{mn} - (12\leftrightarrow 3)\bigr]\cr
& + (k^{1}\cdot k^2)\bigl[ 2 A^1_{[n} (\g_{m]} W^{23})_\a + 2 A^{13}_{[n} (\g_{m]} W^{2})_\a -(1\leftrightarrow 2)\bigr]\cr
& + (k^{12}\cdot k^3)\bigl[  2 A^{12}_{[n} (\g_{m]} W^3)_\a - (12\leftrightarrow 3)\bigr]\,.&\DFthree\cr
}$$
The above equations give rise to a natural rank-three definition of multiparticle SYM equations of
motion: The non-contact terms in \EOMAthree, \EOMAmthree, \EOMWthree\ and \DFthree\ perfectly tie in with those in the two-particle equations of motion \EOMAtwo\ to \EOMFtwo. Note that the contact terms in $D_\a A^m_{123}$ and $D_\a W^{\b}_{123}$ are related via $A^m_C \leftrightarrow W^\a_C$
where $C$ denotes a multiparticle label, see \EOMAmthree\ and \EOMWthree. The additional contact terms of the form $A^{B}_{[n} (\g_{m]} W^C)$ in $D_\a F^{123}_{mn}$ have their two-particle analogues in the second line of \EOMFtwo.

\subsubsec Symmetry properties at rank three

The rank-three superfields defined above are manifestly antisymmetric in the first two labels, so
they satisfy $\lie_2$ from \BRSTsym.
However, one can show using the explicit expressions above that only a subset of the rank-three
superfields satisfies $\lie_3$,
\eqn\JacWthreeFthree{
\lie_3\circ \hat A^{123}_\a \neq 0,\quad\lie_3\circ \hat A^{123}_m \neq 0,\quad\lie_3\circ W^\a_{123} =\lie_3\circ F^{123}_{mn} = 0.
}
This explains the non-hatted notation for $W_{123}^\a$ and $F^{123}_{mn}$; they are BRST blocks already.
To obtain BRST blocks for the other superfields they need to be redefined in order to satisfy the symmetry $\lie_3$.
Fortunately, the underlying system of equations of motion greatly simplifies this task.

To see this,
note that since $\lie_3\circ W_{123}^\a =0$ equation \EOMAmthree\ implies that,
\eqn\QAmVhat{
D_\a\bigl( \lie_3\circ \hat A^{123}_m\bigr)  = k^{123}_m\bigl( \lie_3\circ \hat A_\a^{123}\bigr).
}
And it turns out that $k^{123}_m$ can be factored out in the cyclic sum of $\hat A_m^{123}$,
\eqn\Amtrivial{
\lie_3\circ \hat A^m_{123} = 3\,k_{123}^m H_{123}\,,
}
where
\eqn\Rthree{
H^{123} = {1\over 6}\bigl[(A^1\cdot A^{23}) - (k^2_p - k^3_p)A^p_1 (A^2\cdot A^3) +  {\rm cyclic}(123)\bigr]\,.
}
Therefore the redefinitions
\eqnn\redefsthree
$$\eqalignno{
A^{123}_m &= \hat A^{123}_m - k^{123}_m H^{123}\,,\cr
A_\a^{123} &= \hat A_\a^{123} - D_\a H^{123}\,, &\redefsthree\cr
}$$
imply that $A_\a^{123}$ and $A_m^{123}$ are BRST blocks since,
$$
\lie_2\circ A_\a^{123} = \lie_2\circ  A_m^{123} = \lie_3\circ  A_\a^{123} = \lie_3\circ  A_m^{123} = 0.
$$
This is a significant simplification compared to the
redefinition \Tdef. The latter required an ``inversion'' of the BRST charge
on $\lie_3\circ ( L_{2131}+\ldots)$ whereas \Amtrivial\ extracts the rank-three redefinition $H_{123}$ from
a straightforward $\lie_3$ operation on the known expression \Amthree\ for $\hat A^m_{123}$.

It is easy to show that $F^{123}_{mn}$ from \Fthree\ can now be rewritten as a field-strength
using the BRST block $A_m^{123}$,
\eqnn\FSthree
$$\eqalignno{
F^{123}_{mn} &= k^{123}_m A^{123}_n - k^{123}_n A^{123}_m \cr
&\quad{} - (k^1\cdot k^2)\bigl[ A^1_m A^{23}_n - A^1_n A^{23}_m - (1\leftrightarrow 2)\bigr]\cr
&\quad{} - (k^{12}\cdot k^3)\bigl[ A^{12}_m A^3_n - (12\leftrightarrow 3)\bigr]\,. &\FSthree
}$$
Thus \Fthree\ satisfying the symmetry $\lie_3 \circ F^{123}=0$ can be understood as
a property inherited from $A^m_{123}$ since the contact term structure of \FSthree\ is the
same as in the equation of motion $D_\a A^{123}_m$ from which the BRST symmetry
was derived in the first place.

Defining rank-three vertex operators
\eqn\ThreeVertices{
V_{123} = \l^\a A^{123}_\a,\quad  U^{123} = \p\t^\a A^{123}_\a + \Pi^m  A^{123}_m + d_\a W^\a_{123} + \half N^{mn}F^{123}_{mn}\,,
}
it follows  that \PSEOM\ as well as \QVtwo\ and \QUtwo\ have a rank-three counterpart,
\eqnn\QVthree
\eqnn\QUthree
$$\eqalignno{
QV_{123} &= (k^1\cdot k^2)(V_1 V_{23} + V_{13}V_2) + (k^{12}\cdot k^3)V_{12}V_3, &\QVthree\cr
QU_{123} &= \p V_{123} + (k^1\cdot k^2)\bigl[V_1 U_{23} + V_{13}U_2 - (1\leftrightarrow 2)\bigr]\cr
&\quad{} + (k^{12}\cdot k^3)\bigl[V_{12}U_3 - (12\leftrightarrow 3)\bigr]. &\QUthree
}$$
It is interesting to observe that $\lie_3$ action translates to a total derivative
\eqn\suggestive{
\hat U^{123} + \hat U^{231} + \hat U^{312} = (\p\t^\a D_\a + \Pi^m k^{123}_m) H^{123}= \p H^{123},
}
where $\hat U^{123}$ is related to $U^{123}$ in the obvious way $A^{123}_\a \leftrightarrow \hat A^{123}_\a $ and
$A^{123}_m \leftrightarrow \hat A^{123}_m$. The total worldsheet derivative suggests that the failure of
the $\lie_3$ symmetries in \suggestive\ decouples from string amplitudes and their SYM limit. In view of the
diagrammatic interpretation of $K_{123}$ shown in \figgraft, the vanishing of $U^{123}+U^{231}+U^{312}$ can be
viewed as the kinematic dual of the Jacobi identity $f^{12a} f^{a3b}+f^{23a} f^{a1b}+f^{31a} f^{a2b}=0$ among
color factors. This indicates that the rank three superfields $K_{123}$ of SYM carry the fingerprints of the
BCJ duality between color and kinematics \BCJ.

\subsec Rank four

The patterns from the discussions above suggest how to proceed. The following superfields
\eqnn\Ahatfour
\eqnn\Amhatfour
\eqnn\Whatfour
$$\eqalignno{
\hat A^{1234}_\a &= - \half\Bigl[  A^{123}_\a (k^{123}\cdot  A^4) +  A^{123}_m (\g^m  W^4)_\a - \(123\leftrightarrow 4\)\Bigr]\,,&\Ahatfour\cr
\hat A^{1234}_m &= \half\Bigl[ A^{123}_p  F^4_{pm} -  A^{123}_m (k^{123}\cdot  A^4) 
+ ( W^{123}\g_m  W^4)
- \(123\leftrightarrow 4\)\Bigr] &\Amhatfour\cr
\hat W_{1234}^\a &= {1\over 4}(\g^{rs} W^4)^\a  F^{123}_{rs} - (k^{123}\cdot  A^4) W_{123}^\a - \(123\leftrightarrow 4\)\cr
&\quad{} + \half (k^1\cdot k^2)\Bigl[  W_{23}^\a ( A^1\cdot  A^4) +  W_{2}^\a ( A^{13}\cdot  A^4) - (1\leftrightarrow 2)\Bigr]\cr
&\quad{}+ \half (k^{12}\cdot k^3)\Bigl[ W_3^\a ( A^{12}\cdot  A^4) - (12\leftrightarrow 3)\Bigr] &\Whatfour\cr
%
}$$
manifestly satisfy the $\lie_2$ and $\lie_3$ symmetries of \BRSTsym. In general, by using the fully redefined
BRST-blocks $A^{12\ldots p-1}_\a, A_{12\ldots p-1}^m$ and $W^\a_{12\ldots p-1}$ in the recursive definition of
$A^{12\ldots p}_\a$, there is only one novel Lie symmetry to impose at each rank. This is much more economic compared
to the $p-1$ redefinitions to arrive at $T_{12 \ldots p}$ in \nptTree\ (which additionally required ``inverting'' the
BRST charge and were much more laborious). Once the last Lie symmetry $\lie_4$ is enforced in section 3.3.1, the
recursions \Ahatfour\ to \Whatfour\ for $K_{1234}$ can be given a grafting interpretation similar to rank three, see
\figmoregraft\ and Appendix~A.

\ifig\figmoregraft{Up to $\lie_4$ symmetry redefinitions, the recursions \Ahatfour\ to \Whatfour\ yield rank-four
BRST blocks $K_{1234} $ by combining $K_{123}$
with $K_{4}$. At the level of diagrams, this can be interpreted as grafting the associated trees.}
{\epsfxsize=0.60\hsize\epsfbox{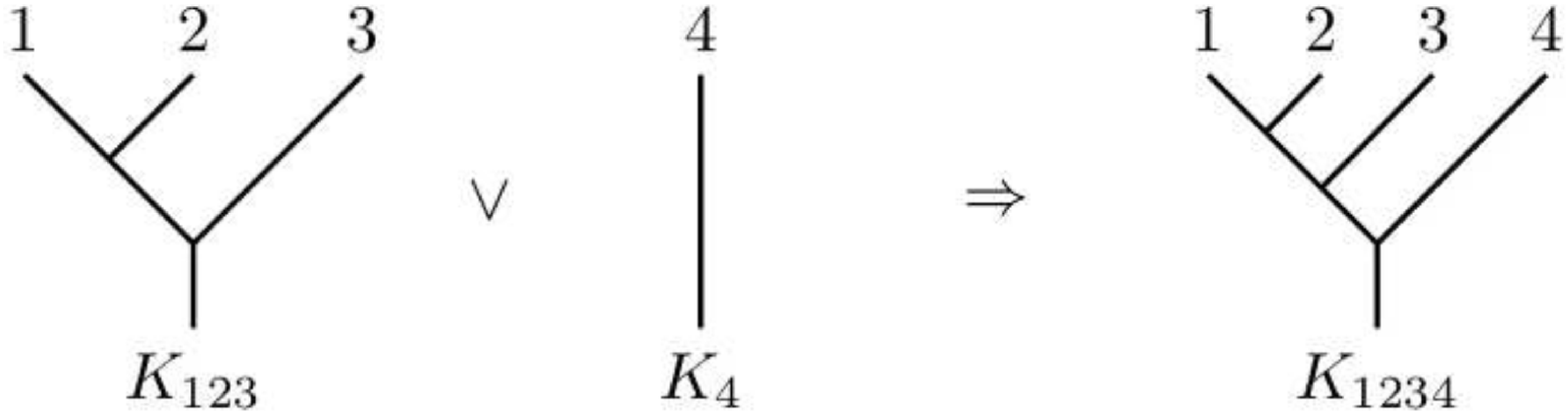}}


The rank-four definitions \Ahatfour\ to \Whatfour\ are guided by the same key principles applied at rank three: repetition of the recursive pattern \Athree, \Amthree\ and \Wthree\ as well as multiparticle equations of motion as in \EOMAthree, \EOMAmthree\ and \EOMWthree. Straightforward but
tedious calculations show that
\eqnn\QAhatfour
\eqnn\QAmhatfour
\eqnn\QWhatfour
$$\eqalignno{
D_\a \hat A^{1234}_\b + D_\b \hat A^{1234}_\a &= \g^m_{\a\b}\hat A_m^{1234}\cr
&\quad{} + (k^1\cdot k^2)\bigl[A^1_\a \hat A^{234}_\b + \hat A^{134}_\a A^2_\b + A^{13}_\a A^{24}_\b  + A^{14}_\a
A^{23}_\b - (1\leftrightarrow 2)\bigr]\cr
&\quad{} + (k^{12}\cdot k^3)\bigl[A^{12}_\a A^{34}_\b  + \hat A^{124}_\a A^3_\b - (12\leftrightarrow 3)\bigr]\cr
&\quad{} + (k^{123}\cdot k^4)\bigl[ A^{123}_\a A^4_\b - (123\leftrightarrow 4) \bigr]&\QAhatfour \cr
D_\a\hat A^{1234}_m &= (\g_m \hat W^{1234})_\a + k^{1234}_m \hat A^{1234}_\a\cr
&\quad{} + (k^1\cdot k^2)\bigl[A^1_\a \hat A^{234}_m + \hat A^{134}_\a A_m^2 + A^{13}_\a A^{24}_m + A^{14}_\a A^{23}_m  - (1\leftrightarrow 2)\bigr]\cr
&\quad{} + (k^{12}\cdot k^3)\bigl[A^{12}_\a A^{34}_m + \hat A^{124}_\a A^3_m -  (12\leftrightarrow 3)\bigr]\cr
&\quad{} + (k^{123}\cdot k^4) \bigl[A^{123}_\a A^4_m  - (123\leftrightarrow 4)\bigr] &\QAmhatfour\cr
D_\a \hat W^\b_{1234} & = {1\over 4}(\g^{mn})_\a{}^\b \hat F^{1234}_{mn}\cr
&\quad{} + (k^1\cdot k^2)\bigl[A^1_\a W_{234}^\b + \hat A^{134}_\a W_2^\b + A^{13}_\a W_{24}^\b + A^{14}_\a W_{23}^\b - (1\leftrightarrow 2)\bigr]\cr
&\quad{} + (k^{12}\cdot k^3)\bigl[A^{12}_\a W_{34}^\b + \hat A^{124}_\a W_3^\b - (12\leftrightarrow 3)\bigr]\cr
&\quad{} + (k^{123}\cdot k^4) \bigl[A^{123}_\a W_4^\b - (123\leftrightarrow 4)\bigr] &\QWhatfour\cr
%
}$$
for some $\hat F^{1234}_{mn}$ whose form is not important at this point.
Note that
the rank-three superfields in the terms proportional to $(k^{123}\cdot k^4)$ are the true BRST blocks
and not their hatted versions.

\subsubsec Symmetry properties at rank four

The hatted superfields appearing in the right-hand side of \QAhatfour\ to \QWhatfour\ can be rewritten in terms of
BRST blocks by using the rank three redefinitions $\hat A^{123}_\a = A^{123}_\a + D_\a H_{123}$
and $\hat A^{123}_m = A_m^{123} + k^{123}_m H_{123}$. The terms containing $H_{ijk}$
can be manipulated to the left-hand side in order to redefine the rank-four superfields. The outcome is,
\eqn\redefprimefour{
K'_{1234} = \hat K_{1234} -
(k^1\cdot k^2)\big(K_2 H_{134} - K_1 H_{234}\big)
- (k^{12}\cdot k^3) H_{124}K_3
}
where $K^{B}$ denotes any of the BRST blocks, $[A^{B}_\a, A^{B}_m, W^\a_{B}]$. For example,
\eqn\Afourredef{
A'^{1234}_\a = \hat A^{1234}_\a -
(k^1\cdot k^2)\bigl(A^2_\a H_{134} - A^1_\a H_{234} \bigr) - (k^{12}\cdot k^3) H_{124}A^3_\a.
}
After the redefinitions of \redefprimefour\ it turns out that the superfield $W'^\a_{1234}$
satisfies all the Lie symmetries \BRSTsym\ up to rank four,
\eqn\Wfoursym{
\lie_2\circ W'^\a_{1234} = \lie_3\circ W'^\a_{1234} = \lie_4\circ W'^\a_{1234} = 0,
}
and therefore $W^\a_{1234}\equiv W'^\a_{1234}$ is a BRST block.

Since $W^\a_{1234}$ satisfies \Wfoursym, it immediately follows from the contact
term structure of \QAmhatfour\ that \QAmVhat\ has the following rank-four analogue
\eqnn\Anothers
$$\eqalignno{
D_\a\bigl(\lie_4\circ A'^{1234}_m\bigr) &= k^{1234}_m \bigl(\lie_4\circ A'^{1234}_\a\bigr)\,. &\Anothers\cr
}$$
Furthermore, a straightforward calculation shows that $k^{1234}_m$ factorizes in
$\lie_4 \circ A'^{1234}_m$,
\eqn\Straight{
\lie_4\circ A'^{1234}_m = 4\,k^{1234}_m H^{1234}\,,
}
and the explicit expression for $H_{1234}$ is displayed in Appendix~\Hfourapp.

Hence, the redefined superfields
\eqnn\redefAmfour
$$\eqalignno{
A^{1234}_m &= A'^{1234}_m - k^{1234}_m H^{1234}&\redefAmfour\cr
A_\a^{1234} &= A'^{1234}_\a - D_\a H^{1234} \cr
}$$
obey the required BRST symmetries:
\eqn\BRSTsymFour{
\eqalign{
\lie_2\circ A^{1234}_\a &= \lie_3\circ A^{1234}_\a = \lie_4\circ A^{1234}_\a = 0,\cr
\lie_2\circ A^{1234}_m &= \lie_3\circ A^{1234}_m = \lie_4\circ A^{1234}_m = 0,
}}
and therefore define rank-four BRST blocks.

Once the expression for $A^m_{1234}$ is known the superfield $F^{1234}_{mn}$ can
be written down immediately in field-strength form,
\eqnn\FSfour
$$\eqalignno{
F^{1234}_{mn} &= k^{1234}_m A^{1234}_n - k^{1234}_n A^{1234}_m\cr
&\quad{} + (k^1\cdot k^2)\bigl[A^1_n A^{234}_{m} +  A^{134}_n A_{m}^2 + A^{13}_n A^{24}_{m} + A^{14}_n A^{23}_{m} - (m\leftrightarrow n)\bigr]\cr
&\quad{} + (k^{12}\cdot k^3)\bigl[A^{12}_n A^{34}_{m} + A^{124}_n A^3_{m} - (m\leftrightarrow n)\bigr]\cr
&\quad{} + (k^{123}\cdot k^4) \bigl[ A^{123}_n A^4_{m} - A^{123}_{m} A^{4}_{n}\bigr]. &\FSfour\cr
}$$
A straightforward but tedious calculation then shows that its expected equation of motion indeed holds,
\eqnn\QFfour
$$\eqalignno{
 D_\a F^{1234}_{mn} &=
 	 k^{1234}_m (\g_{n}  W^{1234})_\a
 	- k^{1234}_n (\g_{m}  W^{1234})_\a\cr
 &\quad{} + (k^1\cdot k^2)\Bigl[ A^1_\a F^{234}_{mn} +  A^{134}_\a F^{2}_{mn} + A^{13}_\a F^{24}_{mn} + A^{14}_\a F^{23}_{mn}  - (1\leftrightarrow 2)\Bigr]\cr
 &\quad{} + (k^{12}\cdot k^3) \Bigl[ A^{12}_\a F^{34}_{mn} +  A^{124}_\a F^{3}_{mn} - (12\leftrightarrow 3)\Bigr]\cr
 &\quad{} + (k^{123}\cdot k^4) \Bigl[  A^{123}_\a F^{4}_{mn} - A^4_\a F^{123}_{mn}\Bigr]\cr
 &\quad{} + (k^1\cdot k^2) \Bigl[ 2A^{1}_{[n} (\g_{m]} W^{234})_\a  + 2A^{134}_{[n} (\g_{m]} W^{2})_\a    \cr
 &\qquad{} + 2A^{14}_{[n} (\g_{m]} W^{23})_\a  + 2A^{13}_{[n} (\g_{m]} W^{24})_\a  - (1\leftrightarrow 2)  \Bigr]\cr
 &\quad{}  + (k^{12}\cdot k^3) \Bigl[ 2 A^{12}_{[n} (\g_{m]} W^{34})_\a  + 2 A^{124}_{[n} (\g_{m]} W^{3})_\a   - (12\leftrightarrow 3)  \Bigr]\cr
 &\quad{}  + (k^{123}\cdot k^4) \Bigl[ 2A^{123}_{[n} (\g_{m]} W^{4})_\a   - (123\leftrightarrow 4)  \Bigr]\,. &\QFfour\cr
}$$
That is why the explicit form of $\hat F^{mn}_{1234}$ was not strictly needed, one can directly write its
BRST-block expression at the end of the redefinition procedure.

Defining rank-four vertex operators
\eqn\UUFour{
V^{1234} = \l^\a A^{1234}_\a, \qquad U^{1234} = \p\t^\a  A^{1234}_\a + \Pi^m  A^{1234}_m + d_\a W^\a_{1234} + \half
N^{mn}F^{1234}_{mn}\,,
} 
it follows that
\eqnn\QVfour
\eqnn\QUfour
$$\eqalignno{
Q V_{1234} &= (k^1\cdot k^2)\bigl[V_1  V_{234} +  V_{134}V_2 + V_{13}V_{24} + V_{14}V_{23}\bigr]\cr
& + (k^{12}\cdot k^3)\bigl[V_{12}V_{34} +  V_{124}V_3\bigr]\cr
& + (k^{123}\cdot k^4)  V_{123}V_4  &\QVfour\cr
QU_{1234} & = \p V_{1234} + (k^1\cdot k^2)\bigl[V_1 U_{234} + V_{13}U_{24} + V_{14}U_{23} + V_{134} U_2 - (1\leftrightarrow 2)\bigr]\cr
& + (k^{12}\cdot k^3)\bigl[V_{12} U_{34} + V_{124}U_{3} - (12\leftrightarrow 3)\bigr]\cr
& + (k^{123}\cdot k^4)\bigl[V_{123} U_{4} - (123\leftrightarrow 4)\bigr].&\QUfour\cr
}$$
And similarly as at rank three, it is interesting that the failure of the $\lie_4$ symmetry to hold for the primed
superfields is equivalent to a total derivative in the integrated vertex $U'^{1234}$ (i.e. $U^{1234}$ with
$A^{1234}_\a \rightarrow A'^{1234}_\a$ and $A^{1234}_m \rightarrow A'^{1234}_m$). Due to the general expectation for
worldsheet derivatives to decouple from string amplitudes, this is another example for the fundamental role played by
Lie symmetries. More specifically, $\lie_4$ compatibility of $U^{1234}$ is a kinematic equivalent of Jacobi identities
among permutations of $f^{12a} f^{a3b} f^{b4c}$. Hence, also the rank four BRST blocks satisfying $\lie_4 \circ
K_{1234}=0 $ point towards the BCJ-duality \BCJ.

\subsec Recursive construction at general rank

Suppose that all the BRST blocks up to rank $p-1$ are known
\eqn\known{
\{A^{12\ldots k}_\a, A^{12\ldots k}_m, W_{12\ldots k}^\a, F^{12\ldots k}_{mn}\}, \qquad 1\le k\le p-1
}
together with the superfields $H_{12\ldots k}$ for $3\le k\le p-1$ used in their construction. The following
steps can be used to obtain the explicit expressions for the rank-$p$ BRST blocks:

\medskip
\item 1. Define a set of rank-$p$ superfields
$\hat K_{12 \ldots p} =\{\hat A^{12\ldots p-1}_\a, \hat A^{12\ldots p-1}_m, \hat W_{12\ldots p-1}^\a\}$
as follows,
\eqnn\General
$$\eqalignno{
\hat A^{12\ldots p}_\a &= - \half\bigl[ A^{12\ldots p-1}_\a (k^{12\ldots p-1}\cdot  A^p)
+ A^{12\ldots p-1}_m (\g^m  W^p)_\a - (12\ldots p-1\leftrightarrow p)\bigr]  \cr
\hat A^{12\ldots p}_m &= \half\bigl[  A^{12\ldots p-1}_n  F^p_{nm}
+  A^{p}_m (k^p\cdot  A^{12{\ldots} p-1}) + ( W^{12{\ldots} p-1}\g_m W^p)
- (12\ldots p-1 \leftrightarrow p)\bigr]\cr
\hat W_{12{\ldots} p}^\a &= {1\over 4}(\g^{rs} W^p)^\a  F^{12{\ldots} p-1}_{rs}
- (k^{12{\ldots} p-1}\cdot A^p) W_{12{\ldots} p-1}^\a  - (12{\ldots} p-1 \leftrightarrow p) &\General\cr
&{}- \sum_{j=2}^{p-1}\sum_{\d \in P(\g_j)} (k^{1\ldots {j-1}}\cdot k^j)\bigl[ W_{1\dots j-1,\{\d\}}^\a (A^{j,\{\g_j\backslash \d\}}\cdot A^p)
- (12\ldots j-1\leftrightarrow j)\bigr]\,,
}$$
where the set $\g_j = \{j+1, \ldots, p-1\}$ contains the $p-j-1$ labels between $j$ and $p$ and $P(\g_j)$ is its
power set.
Note that they manifestly obey all the $\lie_k$ symmetries up to rank $k=p-1$, but not (yet) $\lie_p$.

\noindent One can check that the superfields $\hat K_{12 \ldots p}$ satisfy equations of motion
of the form \GeneralEOM\ whose right-hand
side contains not only lower-rank BRST blocks but also their hatted versions, for example,
\eqnn\rankFiveEx
$$\eqalignno{
2D_{(\a} \hat A^{12345}_{\b)}& = \g^m_{\a\b}\hat A^{12345}_m &\rankFiveEx\cr
& + (k^{1}\cdot k^2)\bigl[
	  A^{1}_\a \hat A^{2345}_\b
          + A^{13}_\a \hat A^{245}_\b
	  + A^{14}_\a \hat A^{235}_\b
	  + A^{15}_\a  A^{234}_\b\cr
& \phantom{+(k^{1}\cdot k^2)}
          + A^{134}_\a A^{25}_\b
	  + \hat A^{135}_\a A^{24}_\b
	  + \hat A^{145}_\a A^{23}_\b
          + \hat A^{1345}_\a  A^{2}_\b
	  - (1\leftrightarrow 2)
	  \bigr]\cr
&	  + (k^{12}\cdot k^3)\bigl[
           A^{12}_\a  \hat A^{345}_\b
          + A^{124}_\a A^{35}_\b
	  + \hat A^{125}_\a A^{34}_\b
          + \hat A^{1245}_\a  A^{3}_\b
          - (12\leftrightarrow 3)
          \bigr]\cr
&       + (k^{123}\cdot k^4)\bigl[
           A^{123}_\a A^{45}_\b
          + \hat A^{1235}_\a  A^{4}_\b
	  - (123\leftrightarrow 4)
          \bigr]\cr
&       + (k^{1234}\cdot k^5)\bigl[ A^{1234}_\a A^{5}_\b - (1234\leftrightarrow 5)\bigr]\,.\cr
}$$
However, they can
be redefined $\hat K_{12 \ldots p}\rightarrow K'_{12 \ldots p}$ such that equations of motion for $K'_{12 \ldots p}$
are written entirely in terms of BRST blocks with rank less than $p$. This leads to the second step:

\medskip
\item 2. Redefine the superfields according to
\eqn\GeneralRedef{
K'_{12 \ldots p} = \hat K_{12 \ldots p} -
\sum_{j=2}^{p-1}\sum_{\d \in P(\g_j)} (k^{1\ldots {j-1}}\cdot k^j)
\bigl[ H_{1\ldots j-1,\{\d\},p}\; K_{j,\{\g_j\backslash \d\}}
- (12\ldots j-1\leftrightarrow j)\bigr]
}
with the constraints $H_{i} = H_{ij} = 0$. For example,
$$\eqalignno{
K'_{12345} &= \hat K_{12345}\cr
& - (k^1\cdot k^2)\bigl[
H_{1345} K_2 + H_{145} K_{23} + H_{135} K_{24} - (1\leftrightarrow 2)\bigr]\cr
& - (k^{12}\cdot k^3)\bigl[
H_{1245} K_3 + H_{125} K_{34} - H_{345} K_{12}\bigr]\cr
& - (k^{123}\cdot k^4)\bigl[ H_{1235}K_4\bigr]\,.
}$$
At this point it turns out that
$W'^\a_{12\ldots p}$ satisfies all the rank-$p$ Lie symmetries, i.e.\
\eqn\LieMiracle{
\lie_k\circ W'^\a_{12\ldots p} = 0,\quad 2\le k\le p.
}
Therefore
$W'^\a_{12\ldots p}\equiv W^\a_{12\ldots p}$ will be
the definition of the spinor field-strength BRST block.

\noindent
As a consequence of \LieMiracle, the following equations will hold,
\eqn\LieConjecture{\eqalign{
D_\a \bigl(\lie_p\circ A'^{12 \ldots p}_m\bigr) &= k^{12 \ldots p}_m {\lie}_p\circ A'^{12 \ldots p}_\a,\cr
{\lie}_p\circ A'^m_{12 \ldots p} &= p\, k^m_{12 \ldots p} H_{12 \ldots p}
}}
where the second equation can be regarded as the definition of $H_{12 \ldots p}$.

\medskip
\item 3. The rank-$p$ BRST blocks are defined as,
\eqnn\GeneralBRSTblocks
$$\eqalignno{
A_\a^{12 \ldots p } &= A'^{12 \ldots p}_\a - D_\a H^{12 \ldots p} &\GeneralBRSTblocks\cr
A^{12 \ldots p}_m &= A'^{12 \ldots p}_m - k^{12 \ldots p}_m H^{12 \ldots p}\cr
W^\a_{12 \ldots p } &= W'^\a_{12 \ldots p}\cr
F^{12 \ldots p}_{mn} &= k^{12 \ldots p}_m A^{12 \ldots p}_n - k^{12 \ldots p}_n A^{12 \ldots p}_m\cr
&+ \sum_{j=2}^{p} \sum_{\d \in P(\beta_j)}  (k_{12 \ldots j-1}\cdot k_j)\,
2 A_{[n}^{12{\ldots} j-1,\{\d\}}\,  A_{m]}^{j, \{\beta_j \backslash \d \} }\,, \cr
}$$
where
the set $\b_j = \{j+1, j+2,{\ldots},p\}$ contains the $p-j$ labels to the right of $j$ and $P(\beta_j)$ denotes its power set.
Note that they satisfy all the Lie symmetries up to rank~$p$.

\medskip
\noindent
It is conjectured that the BRST blocks defined in the three-step procedure above
will satisfy the multiparticle equations of motion,
\eqnn\GeneralEOM
$$\eqalignno{
2D_{(\a}  A^{12\ldots p}_{\b)} &=  \g^m_{\a\b} A^{12\ldots p}_m &\GeneralEOM\cr
&{}+ \sum_{j=2}^{p} \sum_{\d \in P(\beta_j)}  (k_{12 \ldots j-1}\cdot k_j)\,
\bigl[ A_\a^{12{\ldots} j-1,\{\d\}}\,  A_\b^{j, \{\beta_j \backslash \d \} } -  (12\ldots j-1 \leftrightarrow j )\Bigr]\cr
D_\a  A^m_{12\ldots p} &=  (\g^m  W_{12{\ldots} p})_\a + k^m_{12{\ldots} p}  A_\a^{12{\ldots} p}\cr
&{}+ \sum_{j=2}^{p} \sum_{\d \in P(\beta_j)}  (k_{12 \ldots j-1}\cdot k_j)\,
\Bigl[  A_\a^{12{\ldots} j-1,\{\d\}}\,  A^m_{j, \{\beta_j \backslash \d \} }
-  (12\ldots j-1 \leftrightarrow j )\Bigr]\cr
D_\a  W^\b_{12\ldots p} &=  {1\over 4}(\g^{mn})_\a{}^\b  F^{12{\ldots} p}_{mn}\cr
&{}+ \sum_{j=2}^{p} \sum_{\d \in P(\beta_j)}  (k_{12 \ldots j-1}\cdot k_j)\,
\Bigl[ A_\a^{12{\ldots} j-1,\{\d\}}\,  W^\b_{j, \{\beta_j \backslash \d \} }
-  (12\ldots j-1 \leftrightarrow j )\Bigr]\cr
D_\a  F^{mn}_{12\ldots p} &= 2 k^{[m}_{12\ldots p} (\g^{n]}  W_{12{\ldots} p})_\a\cr
&{}+ \sum_{j=2}^{p} \sum_{\d \in P(\beta_j)}  (k_{12 \ldots j-1}\cdot k_j)\,
\Bigl[  A_\a^{12\ldots j-1,\{\d\}}\,  F^{mn}_{j, \{\beta_j \backslash \d \} }\cr
&{}\qquad + 2  A^{[n}_{12\ldots j-1,\{\d\}}\, (\g^{m]}  W_{j, \{\beta_j \backslash \d \} })_\a
-  (12\ldots j-1 \leftrightarrow j )\Bigr]\,.\cr
}$$

Furthermore, defining the multiparticle vertex operators as
\eqn\UUFour{
V^{B} = \l^\a A^{B}_\a, \qquad U^{B} = \p\t^\a  A^{B}_\a + \Pi^m  A^{B}_m + d_\a W^\a_{B} + \half N^{mn}F^{B}_{mn}\,,
}
one can show using the equations of motion \GeneralEOM\ that they satisfy
\eqnn\BRSTvarGeneral
$$\eqalignno{
QV_{12{\dots} p} &= \sum_{j=2}^{p} \sum_{\a \in P(\beta_j)} (k^{12\ldots j-1}\cdot k^j)\,
V_{12\ldots  j-1,\{\a\}}\; V_{j, \{\beta_j \backslash \a\}},&\BRSTvarGeneral\cr
QU_{12{\dots} p} &= \p V_{12 \ldots p} + \sum_{j=2}^{p} \sum_{\a \in P(\beta_j)} (k^{12\ldots j-1}\cdot k^j)
 \bigl[ V_{12\ldots  j-1,\{\a\}}\; U_{j, \{\beta_j \backslash \a\}} - (12 \ldots j-1 \leftrightarrow j)\bigr]\,.
}$$

It is interesting to note that there is an alternative definition\foot{In fact, this is
the representation chosen in all the checks performed with a computer.} of the rank-$p$ BRST blocks $A_\a^{12 \ldots p}$
and $A_m^{12 \ldots p}$ in \GeneralBRSTblocks\ which does not require the explicit knowledge of the rank-$p$ $H_{12 \ldots p}$
(assuming it exists). One can simply project $A'^{12 \ldots p}_\a$
and $A'^{12 \ldots p}_m$ into the kernel of $\lie_p\circ$, for example, use 
${3\over 4}A'^{1234}_m + {1\over 4}\bigl(A'^{1243}_m - A'^{3412}_m + A'^{3421}_m\bigr)$
rather than \redefAmfour\ as a definition for $A_m^{1234}$ and similarly for $A_\a^{1234}$. This is convenient since
it allows to get the complete set of rank $p$ BRST blocks using $H_{12 \ldots k}$ with $k\le p-1$.

We have explicitly constructed BRST blocks up to rank four using the steps above. Furthermore, preliminary checks also
indicate that this construction works for rank five.

\newsec Berends--Giele currents

In the 1980's, Berends and Giele introduced the concept of gluonic tree amplitudes with one off-shell leg and found a recursive
construction for these so-called ``currents'' \BerendsME. Physical amplitudes are easily recovered by removing the off-shell propagator
(as represented by the dots in \figthree) from the current. In the following, we construct ten-dimensional superspace representations
of Berends-Giele currents from multiparticle SYM superfields. The particular combinations of rank-$p$ superfields is firstly guided by
the cubic diagrams of a $p+1$ tree amplitude. Secondly, it turns out that the contact terms of their multiparticle equations
of motion \GeneralEOM\ simplify when following the diagrammatic intuition.

This construction has been partially realized in \nptMethod\ for the superpotential $A^{12 \ldots p}_\a$ which suffices to determine
the SYM tree amplitude from a supersymmetric Berends--Giele recursion. In the superspace setup, the divergent off-shell propagator is cancelled by
the BRST charge, see section 5.1. At one-loop level \oneloopbb, Berends--Giele
 currents from the field strengths $W_{12\ldots p}^\alpha, F_{12\ldots p}^{mn}$ were assembled to BRST-invariant kinematic factors.
 We shall now provide a unified discussion of all the Berends--Giele currents associated with the multiparticle
 superfields of the previous section.

\ifig\figthree{From cubic diagrams $K_A$ to Berends--Giele currents ${\cal K}_A$.}
{\epsfxsize=0.75\hsize\epsfbox{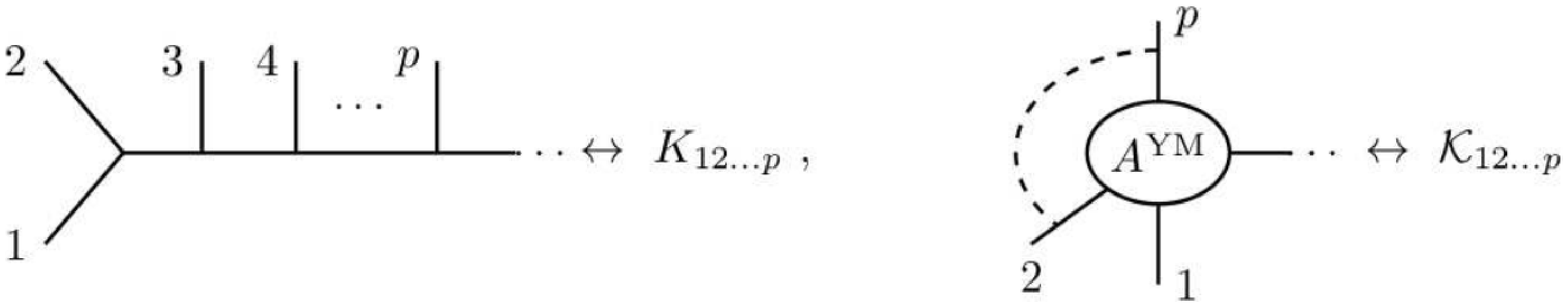}}


For each multiparticle superfield $K_B \in \{ A^{B}_\alpha, A^m_{B} ,
W_{B} ^\alpha, F_{B} ^{mn} \}$ with $B=12\ldots p$ we define a ghost-number zero Berends--Giele current
${\cal K}_B \in \{  \cA^{B}_\alpha, \cA^m_{B} , \cW_{B} ^\alpha, \cF_{B} ^{mn} \}$ as follows: Firstly
decorate the cubic diagrams represented by $K$ with their propagators and secondly
combine the propagator-dressed diagrams such that they resemble a color-ordered
Yang--Mills tree amplitude with an off-shell leg \BerendsME, see \figthree\foot{See Appendix~A.3 for a more
mathematical approach to this diagrammatic construction.}.
As pointed out in \Polylogs, this is implemented through the inverse momentum
kernel \refs{\oldMomKer,\MomKer}\foot{In the conventions of \Polylogs, $S[\s|\rho]_1$ is symmetric under exchange of
$\s$ and $\rho$. For example, the rank two
and three versions of its inverse are given by
$$
S^{-1}[ 2 | 2 ]_1 = {1\over s_{12}} ,\quad S^{-1}[23|23]_1 =  {1\over s_{12} s_{123} }+{1\over s_{123} s_{23} },\quad
S^{-1}[23|32]_1 = - {1\over s_{123} s_{23}}\,.
$$}
\eqn\BRSTc{
{\cal K}_{1\sigma(23\ldots p)} \equiv \sum_{\rho \in S_{p-1}} S^{-1}[\sigma|\rho]_1 \, K_{1\rho(23\ldots p)} \ ,
}
where $\sigma \in S_{p-1}$, and the momentum kernel $S[\cdot | \cdot ]_1$ is defined as
$$
S[ 2_\rho,\ldots,p_\rho |  2_\sigma,\ldots,p_\sigma ]_1 \equiv \prod_{j=2}^{p} \big(  s_{1,j_\rho} + \sum_{k=2}^{j-1} \theta(j_\rho,k_\rho)  s_{j_\rho,k_\rho}  \big)  \ .
$$
We use the shorthands $s_{ij} = k^i\cdot k^j$ and $i_\rho \equiv \rho(i)$, and the object $\theta(j_\rho,k_\rho)$
equals 1 (zero) if the ordering of the legs $j_\rho,k_\rho$ is the same (opposite) in the ordered sets
$\rho(2,\ldots,p)$ and $\sigma(2,\ldots,p)$. In other words, it keeps track of labels which swap their relative
positions in the two permutations $\rho$ and $\sigma$. At rank $r \leq 4$, for example,
\eqnn\BRSTa
$$\displaylines{
{\cal K}_{12} = { K_{12} \over s_{12}}, \quad \quad {\cal K}_{123} = {K_{123} \over s_{12} s_{123}} + { K_{321} \over
s_{23} s_{123}}\,,\hfil\BRSTa\hfilneg\cr
{\cal K}_{1234} = {1 \over s_{1234}} \Big( {K_{1234}\over s_{12}s_{123} }
 + {K_{3214}\over s_{23}s_{123} } + {K_{12[34]}\over s_{12}s_{34} }
 + {K_{3421} \over s_{34}s_{234} } + {K_{3241} \over s_{23}s_{234} } \Big) \ ,
}$$
and \BGfour\ illustrates that the given expression for ${\cal K}_{1234}$ reproduces
the five cubic diagrams in a color-ordered SYM five-point amplitude with an off-shell leg.

\ifig\BGfour{The Berends--Giele current ${\cal K}_{1234}$ of \BRSTa\ is given by
the sum of the superspace expressions associated with the above five cubic graphs with one leg off-shell.
The mapping between the cubic graphs and BRST blocks is introduced in section~3, fig.~2 and explained in more detail
in appendix~A.}
{\epsfxsize=0.95\hsize\epsfbox{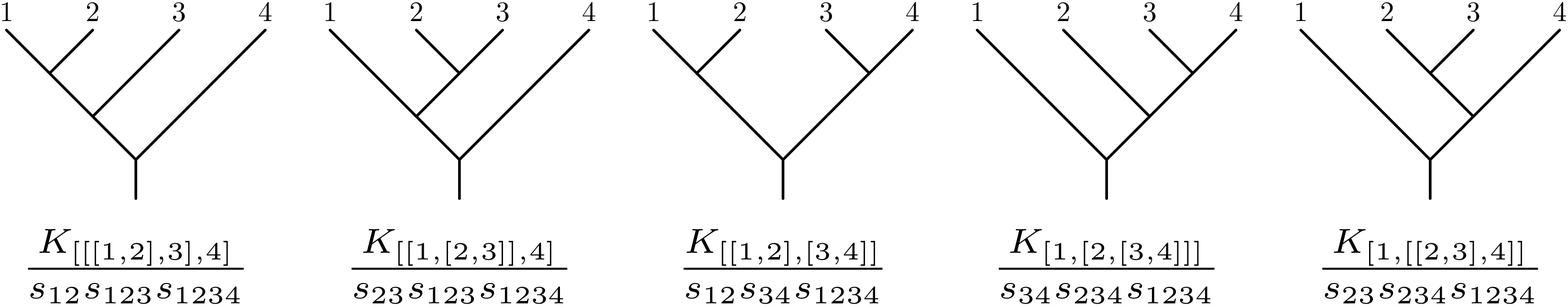}}

The ghost-number zero Berends--Giele currents
${\cal K} \in \{  \cA^{12\ldots p}_\alpha, \cA^m_{12\ldots p} , \cW_{12\ldots p} ^\alpha, \cF_{12\ldots p} ^{mn} \}$
generalize the ghost-number one
analogues $M_{12\ldots p}$ studied in \refs{\nptMethod, \nptTree} which correspond to the unintegrated multiparticle
vertex as
\eqn\VAMAdef{
\cV_A \equiv \lambda^\alpha \cA^{A}_\alpha \equiv M_A \, .
}
One can show using the equations of motion \GeneralEOM\ that the BRST charge acts on Berends--Giele currents of any
ghost number by simple deconcatenation of labels
\eqnn\QBGs
$$\eqalignno{
&QM_{12\ldots p} = \sum^{p-1}_{j=1} M_{12\ldots j}M_{j+1\ldots p} \, , &\QBGs
}$$
as well as
\eqnn\oldQBGs
$$\displaylines{
Q\cA^m_{12\ldots p} = (\lambda \gamma^m \cW_{12\ldots p})
+ k_{12\ldots p}^m \cV_{12\ldots p} + \sum_{j=1}^{p-1}(\cV_{12\ldots j} \cA^m_{j+1\ldots p}
-  \cV_{j+1\ldots p} \cA^m_{12\ldots j}) \hfil\oldQBGs\hfilneg \cr
Q\cW^\alpha_{12\ldots p} ={1\over 4}(\lambda \gamma_{mn})^\alpha \cF^{mn}_{12\ldots p}
+\sum^{p-1}_{j=1} (\cV_{12\ldots j}\cW^\alpha_{j+1\ldots p}-\cV_{j+1\ldots p}\cW^\alpha_{12\ldots j}) \cr
Q\cF^{mn}_{12\ldots p} =2 k^{[m}_{12\ldots p} (\lambda \gamma^{n]} \cW_{12\ldots p})
+ \sum^{p-1}_{j=1} (\cV_{12\ldots j}\cF^{mn}_{j+1\ldots p}-\cV_{j+1\ldots p}\cF^{mn}_{12\ldots j})\cr
\qquad{} + \sum^{p-1}_{j=1} 2\big[ \cA^{[n}_{12\ldots j} (\lambda \gamma^{m]} \cW_{j+1\ldots p})
-\cA_{j+1\ldots p}^{[n} (\lambda \gamma^{m]} \cW_{12\ldots j}) \big] \ .\cr
}$$
By comparing the above equations with \GeneralEOM\ one sees that the kinematic poles in the definition of the
Berends--Giele currents absorb all the explicit kinematic invariants $(k^{12\ldots j-1}\cdot k^j)$ from the right-hand
side of the BRST variations. The extra simplicity of \QBGs\ and \oldQBGs\ compared to \GeneralEOM\ suggests that the
Berends--Giele basis of tree subdiagrams is particularly suitable for a systematic construction of BRST-invariants,
see section~5.

\subsec Symmetries of Berends--Giele currents

Under the momentum kernel multiplication \BRSTc, the Lie-symmetries of the multiparticle
superfields $K_{12 \ldots p}$ are mapped to a different set of Berends--Giele symmetries of ${\cal K}_{12 \ldots p}$,
$$
{\cal K}_{12} + {\cal K}_{21} = 0 \ , \ \ \ \  {\cal K}_{123}  -  {\cal K}_{321}  = {\cal K}_{123} + {\cal K}_{231} + {\cal K}_{312} = 0  , \ \ \ \ \ldots
$$
which leave the same number $(p-1)!$ of independent components at rank $p$. Universality of the
momentum kernel implies that any of the ${\cal K}_{12 \ldots p}$ shares the same symmetry
properties as $M_{12 \ldots p}$ discussed in \refs{\nptMethod,\nptTree}, namely\foot{As a consequence,
we have ${\cal K}_{\a\shuffle\b} = 0$,  $\forall \ \a,\b$.}
\eqn\BGsym{
{\cal K}_{\{\b\},1,\{\a\}} = (-1)^{n_\b} {\cal K}_{1,\a\shuffle\b^T} \ .
}
The notation $\{\b^T\}$ represents the set with the reversed ordering of its $n_\b$ elements and $\shuffle$ denotes
the shuffle product. Furthermore, the convention ${\cal K}_{\ldots \alpha \shuffle \beta\ldots} \equiv \sum_{\sigma\in \a\shuffle\b}
{\cal K}_{\ldots \{\sigma\} \ldots}$ has been used. The multiparticle label $B$ in ${\cal K}_B$ now carries Berends--Giele symmetries \BGsym\ rather than the Lie symmetries \BRSTsym\ of the associated $K_B$.

The symmetry properties \BGsym\ of rank-$p$ currents can be viewed as rank-$(p+1)$
Kleiss--Kuijf relation \KKref\ obeyed by Yang--Mills tree amplitudes where the last leg $p+1$ is off-shell and not displayed,
leaving $(p-1)!$ independent components. Note, however, that the off-shell-ness of one leg in the diagrammatic interpretation
of Berends--Giele currents obstructs an analogue of the BCJ relations \BCJ\ among Yang--Mills tree amplitudes.

On the other hand, an interesting perspective on BCJ relations is opened up when the recursions \General\ for BRST
blocks are rewritten in terms of Berends--Giele currents. This observation is presented in Appendix B, which leads to
a simplified rewriting of one-loop kinematics in terms of SYM amplitudes as compared to \oneloopbb.

\newsec Application to the one loop cohomology

In this section, we explore examples at one-loop how the universal multiparticle equations of motions \GeneralEOM\ and the simplified contact terms in the Berends--Giele picture \QBGs\ and \oldQBGs\ facilitate the construction of BRST invariants. The scalar BRST cohomology at one-loop has been investigated in \oneloopbb\ and identified in the non-anomalous part of open string amplitudes. The trial-and-error construction of the invariants' expansion in terms of Berends--Giele currents is now replaced by a clean recursion. The same mechanisms are applied to novel vectorial invariants which play a key role in closed string amplitudes at one loop, e.g. for S-duality \oneloopMichael\ and for loop momentum dependence in the numerators of the field theory limit \wip.

\subsec Tree level SYM amplitudes

As shown in \nptMethod, tree amplitudes $A^{\rm YM}$ of ten-dimensional SYM theory take an elegant form in
pure spinor superspace,
\eqn\AYM{
A^{\rm YM}(1,2, \ldots,n) = \langle E_{12 \ldots n-1}V_n\rangle\,.
}
The central object $E_{12 \ldots n-1}$ belongs to the BRST cohomology in the momentum phase
space of $n$ massless particles\foot{The restriction on the momentum phase space follows from the fact that the solution $M_{12 \ldots n-1}$ in
$E_{12 \ldots n-1} = QM_{12 \ldots n-1}$ is proportional to a divergent propagator $s^{-1}_{12 \ldots n-1}$.}. Its explicit form can be written in terms of
the Berends--Giele
currents associated with the (generalized) unintegrated vertex
$V_A$ as follows,
\eqn\Edef{
E_{12 \ldots p} = \sum^{p-1}_{j=1} M_{12\ldots j}M_{j+1\ldots p}\,.
}
The pure spinor bracket $\langle \ldots \rangle$ in \AYM\ denotes a zero-mode integration prescription of schematic form $\langle \lambda^3 \theta^5 \rangle=1$. It extracts the
gluon and gluino components of the enclosed superfields 
\psf\ as has been automated in \PSS. The explicit form of the SYM amplitudes in terms of polarization vectors and gaugino wavefunctions up
to multiplicity eight can be downloaded from \WWW.

The BRST cohomology techniques that were used in \nptMethod\ to cast the SYM scattering amplitudes into 
the form \AYM\ also played a crucial role in obtaining  the general solution of the $n$-point tree-level amplitude of
massless open superstrings \nptTree.

\subsec Scalar BRST blocks at one-loop

In \oneloopbb\ the pure spinor zero-mode saturation rules in one-loop amplitudes of the open superstring
were used to obtain an effective prescription to identify contributing
pure spinor superspace expressions: The zero modes of $d_\alpha d_\beta N^{mn}$
extracted from the external vertices are replaced by $(\lambda \gamma^{[m})_\alpha (\lambda \gamma^{n]})_\beta$ . This prescription leads to the
BRST-closed expression $(\l\g^m W^i)(\l\g^n W^j) F^k_{mn}$ in the four-point amplitude \MPS\ and
motivates the following higher-point definitions\foot{$T_{A,B,C}$ and $M_{A,B,C}$ were denoted by $T_A^i T_B^j T_C^k$ and $M_A^i M_B^j M_C^k$ in \oneloopbb, and the representation of $W_A$ and $F_B$ given in the reference is different from the current setup.},
\eqnn\BRSTi
\eqnn\BRSTMi
$$\eqalignno{
T_{A,B,C} &\equiv  {1 \over 3}(\l\g_m W_A)(\l\g_n W_B) F^{mn}_C + (C\leftrightarrow A,B)\,,&\BRSTi\cr
M_{A,B,C} &\equiv  {1 \over 3}(\l\g_m \cW_A)(\l\g_n \cW_B) \cF^{mn}_C + (C\leftrightarrow A,B)\,. &\BRSTMi
}$$
Using the universal form of $Q {\cal W}^\a_{B}$ and $Q {\cal F}^{mn}_B$, one
sees that the BRST variation of \BRSTMi\ is given by deconcatenation of the multiparticle
indices. Regardless of the ranks $|A|, |B|$ and $|C|$, the pure spinor
constraint projects out all terms in \oldQBGs\ with an explicit appearance of $\l^\a$, and we are left with the BRST-covariant expression
\eqn\QMABC{
Q \, M_{A,B,C}  =
\sum_{\ell=1}^{|A|-1} \, \big( \, M_{a_1\ldots a_\ell} \, M_{a_{\ell+1} \ldots a_{|A|},B,C}
-   M_{a_{\ell+1} \ldots a_{|A|}} \, M_{a_1\ldots a_\ell,B,C} \, \big) + (A \leftrightarrow B,C) \ .
}
Note that $QT_{1,2,3} = QM_{1,2,3} =0$ and that $T_{A,B,C}$ and $M_{A,B,C}$ are
totally symmetric in $A$, $B$ and $C$.

\subsec Scalar BRST cohomology at one-loop

The definition \BRSTMi\ of building blocks $M_{A,B,C}$ was used in \oneloopbb\ to construct BRST invariants $C_{1|A,B,C}$ with up to eight particles by trial and error.
We will now present a recursive method to generate them
for arbitrary ranks.

The results of \oneloopbb\ suggest that each term of the form $M_i M_{A,B,C}$, with $i$ a single-particle label,
can be completed to a BRST-closed expression of the schematic form
\eqn\scalarCdef{
C_{i|A,B,C} \equiv M_i M_{A,B,C} + \sum_{ \{\d\}\neq\emptyset}M_{i\{\d\}}f_{\{\d\}} (M_{\cdot,\cdot,\cdot}) \ .
}
As a defining property of the BRST completion for $M_i M_{A,B,C}$, particle $i$ always enters in a multiparticle Berends--Giele current $M_D$. This is formally represented by a sum over (non-empty) ordered subsets $\{\d\}$ of the labels $\{ a_i\}, \{b_i\} , \{c_i\}$ in $A,B,C$ which join particle $i$ in $M_{i\{\d\}}$. The functions $f_{\{\d\}}$ represent the accompanying linear combinations of building blocks $M_{A,B,C}$. 


Nilpotency $Q^2 =0$ implies that $QM_{A,B,C}$ is also BRST closed, and the
form of \QMABC\ suggests that it can be expanded as
\eqnn\QMgen
$$\eqalignno{
Q M_{A , B , C } &=
C_{a_1| a_2\ldots a_{|A|} ,B , C }
- C_{a_{|A|}| a_1\ldots a_{|A|-1} , B , C } \;+\; (A \leftrightarrow B,C)\,.&\QMgen
}$$
We have picked up all the terms $M_i M_{D,E,F}$ in \QMABC\ with single-particle label $i$ and promoted them to BRST completions $C_{i | D,E,F}$. Examples of \QMgen\ can indeed be checked to hold once the explicit expressions for $C_{i|D,E,F}$ are generated. At five points for instance, $C_{1|23,4,5}= M_1 M_{23,4,5} + M_{12} M_{3,4,5}-M_{13} M_{2,4,5}$ (to be derived shortly) allows to verify
\eqnn\exQMgen
$$\eqalignno{
Q M_{123,4,5} &= M_1 M_{23,4,5} + M_{12} M_{3,4,5} - M_{23} M_{1,4,5} - M_3 M_{12,4,5} &\exQMgen\cr
&= C_{1|23,4,5} - C_{3|12,4,5} .
}$$
Now we turn towards the explicit construction of the BRST completion $f_{\{\d\}} (M_{\cdot,\cdot,\cdot})$ in
\scalarCdef. The task is to cancel terms like $M_i( C_{a_{|A|}| a_1\ldots a_{|A|-1} , B , C }-C_{a_1| a_2\ldots
a_{|A|} ,B , C })$ as they appear in $Q(M_i M_{A , B , C }) $ by \QMgen. In order to determine $f_{\{\d\}}
(M_{\cdot,\cdot,\cdot})$ with this property, we define
a linear concatenation operation $\otimes_j$ acting on the multiparticle labels of
Berends--Giele currents $M_A$ as follows,
\eqn\concat{
M_i \otimes_{a_1} M_{a_1 a_2\ldots a_{|A|}}  \equiv M_{i a_1 a_2\ldots a_{|A|}}.
}
In order to ensure that the concatenation $\otimes_{a_1}$ preserves the KK symmetries $M_{a_1 a_2\ldots a_{|A|}}$ of the Berends--Giele currents, we have to specify the leg $a_1$ appearing next to the concatenating label $i$ on the right hand side:
For example, $M_{132}\neq -M_{123}$ implies that $M_1\otimes_3 M_{32} \neq - M_1\otimes_2 M_{23}$ even though $M_{32} = - M_{23}$. The definition \concat\ would be inconsistent with linearity of $\otimes_j$ if the subscript $j$ is unspecified. The $\otimes_j$ action on additional $M_{B,C,D}$ building blocks is defined to be trivial,
$$
M_i \otimes_{a_1} (M_{a_1 a_2\ldots a_{|A|}} M_{B,C,D}) \equiv   (M_i \otimes_{a_1}M_{a_1 a_2\ldots a_{|A|}}) M_{B,C,D} .
$$
As we will see in the following Lemma, there is a neat interplay between action of the BRST charge and the $\otimes_j$ operation defined in \concat.

\proclaim Lemma 1. If $C_{j|A,B,C}$ as defined by \scalarCdef\ is BRST closed, then its concatenation satisfies
\eqn\MBconcat{
Q(M_i\otimes_j C_{j|A,B,C}) = M_i C_{j|A,B,C}.
}
\par
For example, $C_{2|3,4,5} = M_2 M_{3,4,5}$ is BRST closed and $M_1\otimes_2 C_{2|3,4,5} = M_{12}M_{3,4,5}$
satisfies $Q(M_1\otimes_2 C_{2|3,4,5}) = M_{1}M_2M_{3,4,5} = M_1 C_{2|3,4,5}$.

\noindent{\it Proof.} BRST closure of $C_{j|A,B,C}$ amounts to the following ghost number four statement
$$
Q (C_{j|A,B,C}) = \sum_{\{ \sigma \}} M_{j\{\sigma\}}F_{\{\sigma\}} (M_{\cdot}M_{\cdot,\cdot,\cdot}) = 0
$$
with linear combinations $F_{\{\sigma\}}$ of ghost number three objects $M_{\cdot}M_{\cdot,\cdot,\cdot}$. Since $M_{j\{\sigma\}}$ are independent for different sets $\{\sigma\}$, the $F_{\{\sigma\}}$ must vanish individually.
Using the deconcatenation formula \QBGs, one can rewrite the left hand side of \MBconcat\ as follows:
$$\eqalignno{
Q(M_i\otimes_j C_{j|A,B,C}) &=  Q\big( M_{ij}M_{A,B,C} +  \sum_{\{\d\}\neq\emptyset}M_{ij\{\d\}}f_{\{\d\}} (M_{\cdot,\cdot,\cdot}) \big)
\cr
&= M_{i} M_{j}M_{A,B,C} + \sum_{\{\d\}\neq\emptyset} M_i  M_{j\{\d\}}f_{\{\d\}} (M_{\cdot,\cdot,\cdot}) + \sum_{\{ \sigma \}} M_{ij\{\sigma\}}F_{\{\sigma\}} (M_{\cdot}M_{\cdot,\cdot,\cdot})
\cr
&= M_i \Big\{ M_{j}M_{A,B,C} +  \sum_{\{\d\}\neq\emptyset}M_{j\{\d\}}f_{\{\d\}} (M_{\cdot,\cdot,\cdot}) \Big\} \cr
&= M_i C_{j|A,B,C} \ .
}$$
In the first step, we have isolated the first term of $QM_{ij\{\d\}}=M_{i} M_{j\{\d\}}+\ldots$ and the second step
made use of $F_{\{\sigma\}}=0 \ \forall \ \{\sigma\}$ as argued above.\hfill\qed


The following recursive definition can be
checked to generate BRST closed expressions for arbitrary ranks
\eqn\genSol{
C_{i|A,B, C} =
M_iM_{A,B,C} +  \bigl[ M_i \otimes_{a_1} C_{a_1| a_2\ldots a_{|A|} ,B , C }
- M_i \otimes_{a_{|A|}} C_{a_{|A|}| a_1\ldots a_{{|A|}-1} ,B , C }
+ (A \leftrightarrow B,C)\bigr] \ .
}
$Q$-invariance follows from \QMgen\ and Lemma~1 (using the definition $C_{1|\emptyset,A,B}=0$ for single-particle slots). The are $7-2k$ terms in \genSol\ where $k$ is the number of single-particle slots among $A,B,C$. Since $M_i \otimes_{j}$ increases the multiplicity of $C_{j|D,E,F}$ on the right hand side by one, we can regard \genSol\ as a recursion in $|A|+|B|+|C|$. Its first applications up to multiplicity $1+|A|+|B|+|C|=6$ are listed below
\eqnn\BRSTk
$$\eqalignno{
C_{1|2,3,4} &\equiv M_1 M_{2,3,4}&\BRSTk\cr
C_{1|23,4,5} &\equiv M_1 M_{23,4,5} +   M_1\! \otimes_2 \! C_{2|3,4,5} - M_1 \! \otimes_3 \! C_{3|2,4,5}\cr
&= M_1 M_{23,4,5} + M_{12}M_{3,4,5} - M_{13}M_{2,4,5}\cr
C_{1|234,5,6} &\equiv M_{1}M_{234,5,6} + M_1 \! \otimes_2 \! C_{2|34,5,6} - M_1 \! \otimes_4 \! C_{4|23,5,6}\cr
&= M_1 M_{234,5,6} + M_{12}M_{34,5,6} + M_{123}M_{4,5,6} - M_{124}M_{3,5,6}\cr
&\quad{}- M_{14}M_{23,5,6} - M_{142}M_{3,5,6} + M_{143}M_{2,5,6}\cr
C_{1|23,45,6} &\equiv M_1M_{23,45,6} + M_1\! \otimes_2 \! C_{2|45,3,6} - M_1 \!\otimes_3 \! C_{3|45,2,6}
+ M_1 \! \otimes_4 \! C_{4|23,5,6} -M_1 \! \otimes_5 \! C_{5|23,4,6}\cr
&= M_1 M_{23,45,6} + M_{12}M_{45,3,6} - M_{13}M_{45,2,6} + M_{14}M_{23,5,6} - M_{15}M_{23,4,6}\cr
&\quad{}+M_{124}M_{3,5,6} - M_{134}M_{2,5,6}+ M_{142}M_{3,5,6} - M_{152}M_{3,4,6}\cr
&\quad{}-M_{125}M_{3,4,6} + M_{135}M_{2,4,6} - M_{143}M_{2,5,6} + M_{153}M_{2,4,6} \ ,
}$$
and higher-rank expressions are easily obtained as well. Even though the number of terms in $C_{1|234,5,6}$ and $C_{1|23,45,6} $ can be reduced by virtue of the Berends--Giele symmetry $M_{124}+M_{142}=-M_{214}$, we keep the expression in the form $M_{1\ldots}$ compatible with further recursion steps \genSol.

As detailed in 
Appendix B, the $C_{1|A,B,C}$ boil down to linear combinations of SYM tree amplitudes \oneloopbb. Nevertheless, their component expansion up to multiplicity seven can be downloaded from \WWW.


\subsec Vector BRST  blocks at one-loop

In the five-point closed string computation of
\oneloopMichael\ the zero mode saturation in the left/right-mixing sector where the b-ghost contributes
$\Pi^m d_\a d_\b$ led to the definition
\eqn\Wmdef{
W^m_{2,3,4,5} \equiv {1\over 12}(\l\g^n W^2)(\l\g^p W^3)(W^4\g^{mnp}W^5) + (2,3\mid 2,3,4,5),
}
which satisfies
\eqn\QWm{
QW^m_{2,3,4,5} = -(\l\g^m W_2)T_{3,4,5} - (2\leftrightarrow 3,4,5).
}
The notation $(i_1,i_2\mid i_1,{\ldots} ,i_n)$ means a sum over all possible ways of choosing two
indices $i_1$ and $i_2$ out of $i_1,{\ldots} ,i_n$, for a total of ${n\choose 2}$ terms.
Furthermore, another
type of left/right-mixing zero-mode saturation was possible which
required taking $\Pi^m d_\alpha d_\beta N_{np}$ from the integrated vertex
operators, leading to terms of the form $A_2^m T_{3,4,5}$.
The key observation in \oneloopMichael\ was that the vectorial superfield
\eqn\Tmdef{
T^m_{2,3,4,5}\equiv A_2^m\, T_{3,4,5} + (2 \leftrightarrow 3,4,5 ) + W^m_{2,3,4,5}
}
has a BRST variation in which the vector index is carried only by momenta
\eqn\BRSTl{
Q T^m_{2,3,4,5} = k_2^m V_2 T_{3,4,5} + (2 \leftrightarrow 3,4,5 )\,.
}
This fact played a crucial role in demonstrating BRST invariance of the closed-string five-point amplitude \oneloopMichael\
because it allows the BRST variation of the terms contracting left-
and right-movers to factorize and cancel the variation of the holomorphic squared terms.

To generalize this construction to higher multiplicity one defines
\eqnn\GenTmdef
$$\eqalignno{
W^m_{A,B,C,D} &\equiv {1\over 12}(\l\g^n W_A)(\l\g^p W_B)(W_C\g^{mnp}W_D) + (A,B|A,B,C,D)\cr
T^m_{A,B,C,D} &\equiv A^m_A T_{B,C,D} + (A \leftrightarrow B,C,D) + W^m_{A,B,C,D}\,&\GenTmdef\cr
}$$
with multiparticle labels $A,B,C,D$ as well as their Berends--Giele counterparts,
\eqnn\BRSTm
$$\eqalignno{
\cW^m_{A,B,C,D}& \equiv {1\over 12}(\l\g^n \cW_A)(\l\g^p \cW_B)(\cW_C\g^{mnp}\cW_D) + (A,B|A,B,C,D)\cr
M^m_{A,B,C,D} &\equiv {\cal A}^m_A M_{B,C,D} + (A \leftrightarrow B,C,D) + \cW^m_{A,B,C,D}\,,&\BRSTm\cr
}$$
which are totally symmetric in $A,B,C,D$.
The BRST variations \oldQBGs\ -- in particular the universality of the non-contact terms to arbitrary $A$, $B$, $C$ and $D$ -- imply that
\eqnn\QcalW
\eqnn\BRSTn
$$\displaylines{
Q \cW^m_{A,B,C,D} = - (\l\g^m\cW_A) M_{B,C,D} \hfil\QcalW\hfilneg\cr
+ \sum_{j=1}^{|A|-1}\bigl(
M_{a_1 \ldots a_j}\cW^m_{a_{j+1} \ldots a_{|A|},B,C,D}
- M_{a_{j+1} \ldots a_{|A|}}\cW^m_{a_{1} \ldots a_{j},B,C,D}
\bigr)
+ (A\leftrightarrow B,C,D)\cr
QM^m_{A,B,C,D} = k^m_{A} M_A M_{B,C,D}  \hfil\BRSTn\hfilneg \cr
+  \sum_{j=1}^{|A|-1} (M_{a_1 \ldots a_j} M^m_{a_{j+1}\ldots a_{|A|},B,C,D}
- M_{a_{j+1}\ldots a_{|A|}} M^m_{a_1 \ldots a_j,B,C,D})
+(A \leftrightarrow B,C,D)\,.\cr
}$$
The vectorial building block $M^m_{A,B,C,D}$ causes the first explicit appearance of multiparticle vector superfield $A_B^m$, see \BRSTm. Its multiparticle equation of motion in \oldQBGs\ is required to derive \BRSTn\ at arbitrary multiplicities $|A|,\ldots,|D|$. With $M_B = \lambda^\a {\cal A}_\a^B$ and the $W^\a_B, F_B^{mn}$ constituents in the definition \BRSTMi\ of $M_{A,C,D}$, we have by now seen all the four superfields $\{ A_\a^B, A_B^m ,W^\a_B, F_B^{mn}\}$ in the multiparticle vertex operator $U_B$ entering one-loop BRST blocks.

\subsec Vector BRST cohomology at one-loop

It is interesting to study vectorial uplifts $M_{A}M_{B,C,D}\rightarrow M_A M^m_{B,C,D,i}$ of the scalar BRST invariants $C_{1|A,B,C}$ as given by \BRSTk.
The deconcatenation terms due to the second line of \BRSTn\ drop out from the BRST variation, but the contributions from the first line remain where the free vector index is carried by external momenta $k^m$. The first example
\eqn\QBmfive{
B^m_{1|2,3,4,5} \equiv M_1 M^m_{2,3,4,5} , \ \ \ Q B^m_{1|2,3,4,5} = - \big[ k_2^m E_{12} M_{3,4,5}    +
(2\leftrightarrow 3,4,5)\bigr]
}
obtained from $C_{1|2,3,4}$ appeared in the context of the five point closed string amplitude \oneloopMichael. Its six point generalization
$$
B^m_{1|23,4,5,6}\equiv M_1 M^m_{23,4,5,6} + M_1\otimes_2 B^m_{2|3,4,5,6} -  M_1\otimes_3 B^m_{3|2,4,5,6}
$$
resembles $C_{1|23,4,5}$ and satisfies,
\eqn\QBmsix{
QB^m_{1|23,4,5,6} = - k_{2}^m E_{132}M_{4,5,6} + k_3^m E_{123}M_{4,5,6} + \bigl[ k_4^m V_4 C_{1|23,5,6}   +
(4\leftrightarrow 5,6)\bigr]\,.
}
The higher-multiplicity examples are similarly analysed. The fact that the $k^m_i$ coefficients in both \QBmfive\ and \QBmsix\ are $Q$-exact\foot{Recall that $E_{12\ldots p}=Q M_{12\ldots p}$ and $V_4 C_{1|23,5,6} = Q(M_4 \otimes_1 C_{1|23,5,6})$ by Lemma~1.} hints the existence of vectorial BRST invariants.

Vector BRST invariants can be constructed using the same procedures as in the scalar case.
We assume that each superspace expression $M_i M_{A,B,C,D}^m$ with single-particle label $i$ admits a BRST-invariant completion of the form
\eqn\Cmdef{
C^m_{i|A,B,C,D} \equiv M_i M^m_{A,B,C,D} + \sum_{\{\d\}\neq\emptyset}M_{i\{\d\}}f^m_{\{\d\}} (k^m_{\cdot} M_{\cdot,\cdot,\cdot},M^m_{\cdot,\cdot,\cdot,\cdot}) \ .
}
Any term in the sum over ordered subsets $\{\d\}$ of $A\cup B \cup C \cup D$ incorporates label $i$ in a multiparticle $M_{i\{\d\}}$. The accompanying $f^m_{\{\d\}}$ denote vector combinations of building blocks $M^m_{E,F,G,H}$ (see \BRSTm) and $k_H^m M_{E,F,G}$.

Then, as already argued in the scalar case, $Q^2=0$ and the assumed uniqueness of the BRST completions \scalarCdef\ and \Cmdef\ implies that the BRST variation \BRSTn\ can be rewritten as
\eqn\QMnrew{
Q  M^m_{A,B,C,D} = C^m_{a_1|a_2\ldots a_{|A|},B,C,D} -  C^m_{a_{|A|}|a_1 \ldots a_{|A|-1},B,C,D}
+ \delta_{|A|,1} k_{a_1}^m C_{a_1|B,C,D} + (A \leftrightarrow B,C,D) .
}
In the single-particle case $|A|=1$, the first line of \BRSTn\ generates the defining term $M_i M_{B,C,D}$ of a scalar invariant \scalarCdef, and the definition $C^m_{i|\emptyset,B,C,D}\equiv 0$ must then be used to suppress the first two terms of \QMnrew. We take advantage of \QMnrew\ to rewrite $Q (M_i M^m_{A,B,C,D} )$ in terms of $M_i C^m_{j|B,C,D,E}$ and $M_i  C_{j|B,C,D}k_E^m$. Hence, the BRST completions $f^m_{\{\d\}}$ in \Cmdef\ are determined by the BRST ancestors of $M_i  C_{j|B,C,D}$ and $M_i  C^m_{j|B,C,D,E}$. The former are already known from Lemma~1, and the latter can be easily found by the same properties of the concatenation operation \concat. Similar to the scalars $M_{B,C,D}$, the $\otimes_j$ action on vector BRST blocks is defined to be trivial,
$$
M_i \otimes_{a_1} (M^m_{a_1 a_2\ldots a_{|A|}} M_{B,C,D,E}) \equiv   (M_i \otimes_{a_1}M_{a_1 a_2\ldots a_{|A|}}) M^m_{B,C,D,E} .
$$

\proclaim Lemma 2. If $C^m_{j|A,B,C,D}$ as defined by \Cmdef\ is BRST closed, then its concatenation satisfies
\eqn\MBconcat{
Q(M_i\otimes_j C^m_{j|A,B,C,D}) = M_i C^m_{j|A,B,C,D}.
}\par
\noindent{\sl Proof.} The arguments used in the proof of Lemma~1 can be repeated for vectorial
combinations $f^m_{\{\d\}}$ of $k^m_{\cdot} M_{\cdot,\cdot,\cdot}$ and $M^m_{\cdot,\cdot,\cdot,\cdot}$ at ghost number two as well as
$$
Q (C^m_{j|A,B,C,D}) = \sum_{\{ \sigma \}} M_{j\{\sigma\}}F^m_{\{\sigma\}} (k^m_{\cdot}M_{\cdot}M_{\cdot,\cdot,\cdot},M_{\cdot} M^m_{\cdot,\cdot,\cdot,\cdot}) = 0 \ .
$$
The ghost-number-three objects $F^m_{\{\sigma \}}$ built from $k^m_{\cdot} M_{\cdot} M_{\cdot,\cdot,\cdot}$ and
$M_{\cdot} M^m_{\cdot,\cdot,\cdot,\cdot}$ again vanish by independence of the $M_{j\{\sigma\}}$ such that
$$\eqalignno{
Q(M_i\otimes_j C^m_{j|A,B,C,D}) &=   M_{i} \Big\{ M_{j}M^m_{A,B,C,D} + \sum_{\{\d\}\neq\emptyset}  M_{j\{\d\}}f^m_{\{\d\}}   \Big\} + \sum_{\{ \sigma \}} M_{ij\{\sigma\}}F^m_{\{\sigma\}} 
\cr
&= M_i C^m_{j|A,B,C,D}
}$$
by \Cmdef.\hfill\qed

Then, again in analogy with the scalar case,
a recursive definition of vector invariants can be obtained from \QMnrew\ as follows,
\eqnn\veca
$$\eqalignno{
&C^m_{i|A,B,C,D} \equiv M_i M^m_{A,B,C,D} + \big[ \d_{|A|,1} k_{a_1}^m  M_i \otimes_{a_1} C_{a_1|B,C,D} &\veca \cr 
& \ \ +M_i \otimes_{a_1} C^m_{a_1|a_2\ldots a_{|A|},B,C,D} - M_i \otimes_{a_{|A|}} C^m_{a_{|A|}|a_1 \ldots a_{|A|-1},B,C,D}
  + (A \leftrightarrow B,C,D) \, \big] \ . \cr
}$$
BRST invariance follows from \QMnrew\ and Lemma~2. In view of the four slots $A,B,C,D$, the bracket $[\ldots]$ on the right hand side of \veca\ contains $8-n$ terms
where $n$ is the number of single-particle slots.

The first non-trivial applications of \veca\ are easily checked to be BRST closed,
\eqnn\vecb
$$\eqalignno{
C^m_{1|2,3,4,5} &= M_1 M^m_{2,3,4,5} +  \bigl[ k_2^m M_1 \otimes_2 C_{2|3,4,5} + (2\leftrightarrow 3,4,5)\bigr]\cr
&= M_1 M^m_{2,3,4,5} + \bigl[k^2_m M_{12} M_{3,4,5} + (2\leftrightarrow 3,4,5)\bigr]&\vecb \cr
C^m_{1|23,4,5,6} &= M_1 M^m_{23,4,5,6} + M_1 \otimes_2 C^m_{2|3,4,5,6} - M_1 \otimes_3 C^m_{3|2,4,5,6} \cr
&\quad{} +  \bigl[ k_4^m M_1 \otimes_4  C_{4|23,5,6} + (4\leftrightarrow 5,6)\bigr] \cr
&= M_1 M^m_{23,4,5,6} + M_{12} M^m_{3,4,5,6} - M_{13} M^m_{2,4,5,6}\cr
&\quad{}+\big[ k^m_3 M_{123}M_{4,5,6} + (3\leftrightarrow 4,5,6)\bigr]
- \big[ k^m_2 M_{132}M_{4,5,6} + (2\leftrightarrow 4,5,6)\bigr]\cr
&\quad{}+\big[ k^m_4 M_{14}M_{23,5,6} + M_{142}M_{3,5,6} - M_{143}M_{2,5,6} + (4\leftrightarrow 5,6)\bigr]\,,\cr
}$$
and higher-multiplicity analogues are also straightforward to obtain. Component expansion up to multiplicity seven are available from \WWW.

\newsec Conclusion and outlook

In this work, we have constructed multiparticle vertex operators $U^{12\ldots p}$ through a recursive prescription
described in subsection 3.4. This generalizes and streamlines the earlier construction of BRST-covariant building
blocks in \refs{\nptMethod, \nptTree}. The coefficients of conformal weight-one fields $\{\p\t^\a , \Pi^m , d_\a ,N^{mn}\}$
in $U_B$ are interpreted as multiparticle superfields $K_B \in \{A^B_\a, A^B_m , W_B^\a ,F^B_{mn} \}$ of
ten-dimensional SYM with shorthands $B={12\ldots p}$ for external $p$-particle trees. Their equations of motions are
shown to have the same structure as their
single-particle relatives -- see \GeneralEOM\ versus \RankOneEOM. In addition, they are enriched by contact terms where the multiparticle
label $B$ is distributed into two smaller subsets.

These multiparticle SYM fields furnish a kinematic
analogue of the structure constants $f^{abc}$ of the color sector, and their Lie symmetries \BRSTsym\ guarantee that
the tree-level subgraphs described by $K_B$ are compatible with the BCJ duality between color and kinematics \BCJ.
Since the BCJ duality has been observed to hold in various dimensions, it will be interesting to explore lower-dimensional setups for multiparticle equations of motion.

It is worth emphasizing that the Lie-algebraic nature of the BRST blocks is completely general and can be understood
in terms of its basic SYM superfield constituents. The particular combinations of single-particle superfields constituting their
multiparticle generalizations defined in this paper are suggested
by OPE computations among vertex operators in the pure spinor formalism. Moreover, they are in lines with the BRST cohomology organization
of scattering amplitudes
suggested in \towards\ and brought to fruition in \refs{\nptMethod,\nptTree,\oneloopbb}.
Given the general Lie symmetries obeyed by the multiparticle
SYM superfields and their appearance in the OPEs of vertex operators, it is therefore natural to suspect that
the BCJ duality between color and kinematics might be valid at the level of external tree subdiagrams to all
loop-orders \BCJloop.

In section~5, which is devoted to one-loop applications, the zero mode saturation of the minimal pure spinor formalism \MPS\ singles out some elementary
combinations of $K_B$ with beneficial BRST properties -- such as scalars $M_{A,B,C}$ in \BRSTMi\ and vectors $M^m_{A,B,C,D}$ in \BRSTm. We have
derived recursions \genSol\ and \veca\ to construct scalar and vectorial cohomology elements at arbitrary multiplicity
out of $M_{D}M_{A,B,C}$ and $M_E M^m_{A,B,C,D}$. We can learn from the five-point
results in \refs{\oneloopMichael, \wip} that vector invariants are crucial for one-loop amplitudes among closed string
states, where cross-contractions between left- and right-moving worldsheet fields occur.

Since the number of left-right contractions is unbounded for multiparticle one-loop amplitudes, the need for BRST
invariants extends to tensors of arbitrary rank. The construction of tensorial BRST-blocks generalizing $M_{A,B,C}$
and $M^m_{A,B,C,D}$ as well as their BRST-invariant embedding into full-fledged closed string amplitudes is left for
future work \wipH. Moreover, it remains to clarify how
these tensors are related to the gauge anomaly of open superstring amplitudes and its cancellation \GSanomaly.

For all of the aforementioned building blocks, the superspace representation in terms of elementary SYM superfields is
explicitly accessible from this work. So the zero mode integration prescription of the schematic form $\langle
\lambda^3 \theta^5 \rangle =1$ \psf\ as automated in \PSS\ allows to derive supermultiplet
components in terms of gluon polarization vectors and gaugino wave functions. The gluon components of all
the scalar and vector cohomology elements up to multiplicity seven can be found on the website \WWW.

Finally, it is worthwhile to note that the (non-minimal) pure spinor formalism can be interpreted as a critical
topological string \NMPS.
As shown in \lian, the BRST cohomology of a topological CFT is endowed with a Gerstenhaber algebra structure and it
would therefore be interesting to investigate possible connections with the BRST covariance property of
multiparticle vertex operators. As pointed out by in \moore, the associated Gerstenhaber bracket among vertex operators
is a promising starting point to relate string amplitudes of different particle content. These references motivate
further study of multiparticle vertex operators in view of both mathematical structures and applications to scattering
of massive string states.

\bigskip
\noindent{\bf Acknowledgements:} We thank Tim Adamo for useful comments on the draft.
CRM and OS
acknowledge financial support by the European Research Council Advanced Grant No. 247252 of Michael Green. OS is grateful to
DAMTP for hospitality during various stages of this work.

\appendix{A}{Physics of BRST blocks versus Mathematics of cubic graphs}
\applab\appBRSTnotation

\noindent In this Appendix we connect the recursive construction of BRST blocks with mathematical operations on planar
binary trees, see \refs{\vallette,\mathA,\loday} and references therein.
As explained in the references, a mapping between planar binary trees and iterated brackets gives rise to an
explicit Lie algebra basis construction. This will be used to manifest the Lie symmetries \BRSTsym\ of the BRST blocks and
emphasize their connection with cubic graphs which play a central role for the duality between color and kinematics
\BCJ.

\subsec Iterated bracket notation

The antisymmetry of a rank-two BRST block $K_{a_1a_2}$ can be made manifest with the notation
$K_{[a_1,a_2]} \equiv K_{a_1a_2}$.
In general, the defining property of a rank-$p$ BRST block to satisfy all Lie symmetries $\lie_k$ with $k\le p$
motivates the following notation with iterated brackets,
\eqnn\nested
$$\eqalignno{
K_{[a_1,a_2]} &\equiv K_{a_1a_2} &\nested\cr
K_{[[a_1,a_2],a_3]} &\equiv K_{[a_1a_2,a_3]} \equiv K_{a_1a_2a_3}\cr
 &\;\; \vdots \cr
K_{[[[ \ldots[[a_1,a_2],a_3], \ldots ],a_{p-1}],a_p]} &\equiv K_{[a_1a_2 \ldots a_{p-1},a_p]} \equiv
K_{a_1a_2 \ldots a_p}\,.
}$$
The virtue of this bracket structure for the duality between color and kinematics was already emphasized in \dennen.
The above notation reminds of the recursive definition of BRST blocks which
features a repeated antisymmetrization $(a_1a_2 \ldots a_{j-1} \leftrightarrow a_j)$ with $j=2,3, \ldots,p$. Moreover,
they are in lines with the symmetry matching \newstruCte\ with color factors upon expanding the structure constants
\eqn\Tstring{
K_{[[[ \ldots[[a_1,a_2],a_3], \ldots ],a_{p-1}],a_p]} \leftrightarrow {\rm tr}\( [[[
\ldots[[T^{a_1},T^{a_2}],T^{a_3}], \ldots ],T^{a_{p-1}}],T^{a_p}]\)\,.
}
Furthermore, more general bracketing patterns can always be brought to the canonical
form \nested\ by using the antisymmetry and Jacobi identity satisfied by the brackets. For example,
\eqnn\exampLie
$$\eqalignno{
K_{[[1,[2,3]],4]} &= - K_{[[[2,3],1],4]} = - K_{2314} &\exampLie\cr
K_{[[1,2],[3,4]]} &= K_{[[[1,2],3],4]} - K_{[[[1,2],4],3]} =
K_{1234} - K_{1243}\,.
}$$

Using the iterated bracket notation introduced above the explicit expressions for the Lie
symmetries \BRSTsym\ can be easily reproduced. To see this one uses the antisymmetry of the outer commutator to write $K_{[A,B]} = - K_{[B,A]}$
(here $A$ and $B$ represent arbitrary combinations of brackets acting on the multiparticle labels) and applies the
conventions \nested. For example, the  $\lie_4$ symmetry in \BRSTsym\ is reproduced by $K_{[[1,2],[3,4]]} = -
K_{[[3,4],[1,2]]}$, which implies that $K_{1234} - K_{1243} = -
K_{3412} + K_{3421}$.

\subsec Diagrammatic representation of BRST blocks and their recursion

In the mathematics literature, such as \refs{\vallette,\mathA,\loday} and references therein, there is a well-known mapping between planar
binary trees\foot{The precise definitions can be found
in \refs{\vallette,\mathA}. But for our purposes, a planar binary tree is nothing more than a cubic graph with one leg off-shell.} and iterated
brackets which is used to construct an explicit Lie algebra basis \mathA.
Given the iterated bracket convention discussed above, this can be immediately borrowed to create a mapping between cubic
graphs with one leg off-shell and BRST blocks\foot{This prescription
was already hinted (up to an overall sign) in the diagrammatic derivation of the symmetries obeyed by the building block $T_B$ discussed in
\nptTree. The mapping now extends to the whole class of multiparticle superfields $K_B \in
\{A^B_\a, A^m_B, W^\a_B, F^{mn}_B\}$.}, see \binaryTrees.
The algorithm is as follows. First index the external legs with the labels $\{1,2, \ldots, n\}$ from left to right
and, starting from the left, for each vertex associate the bracket $[A,B]$ where $A$ and $B$ represent the labels to
the left and to the right of the vertex (which may already be partially bracketed themselves).

\ifig\binaryTrees{Examples of the mapping between cubic graphs with one leg off-shell and BRST blocks. Together
with the conventions \nested,
the fact that the BRST blocks furnish an explicit representation of the ``Jacobi identity of trees'' of the
type discussed in \BCJ\ becomes manifest.}
{\epsfxsize=0.80\hsize\epsfbox{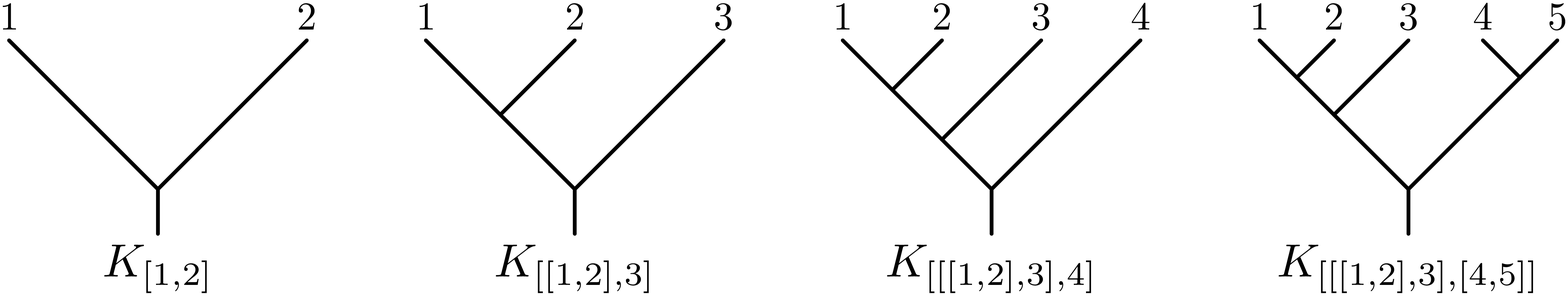}}


Given the mapping described above, it is interesting to consider the effect of the {\it grafting}
\refs{\vallette,\loday} operation of trees in their associated BRST block images. The grafting of two planar
binary trees $t_A$ and $t_B$ is represented by $t_A \vee t_B$ and joins the roots (i.e. the off-shell leg) of $t_A$ and $t_B$ to create a new root.
It is not difficult to see that
if $K_A$ and $K_B$ are the BRST blocks associated with $t_A$ and $t_B$ then $t_A\vee t_B$ is mapped to
$K_{[A,B]}$, see \figgraftApp. Note that the definition of $\hat A_\a^{123 \ldots p}$ in section \BRSTsec\ can be
interpreted (up to the redefinitions by $H_{12 \ldots p}$) as the grafting of two trees with multiplicity $p-1$ and $1$. 

\ifig\figgraftApp{The grafting operation on trees and its corresponding mapping in terms of BRST blocks.}
{\epsfxsize=0.70\hsize\epsfbox{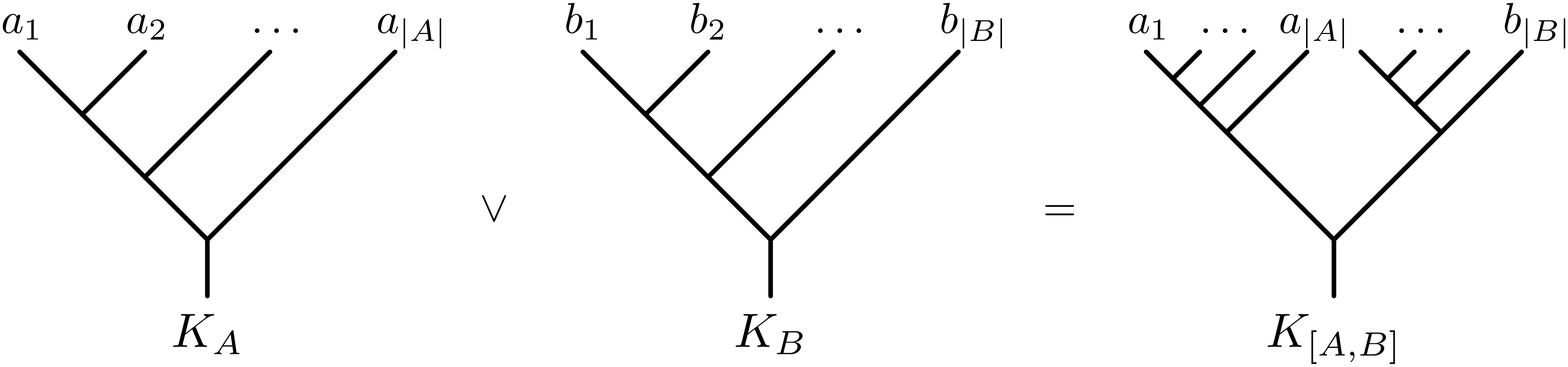}}


\subsec Diagrammatic construction of Berends--Giele currents

It is possible to find the explicit expressions of Berends--Giele currents ${\cal K}_B$ in terms of BRST blocks $K_B$
with a diagrammatic prescription which uses the mapping discussed above. This can be
used as an alternative to the inverse momentum kernel formula given in \BRSTc.

The Berends--Giele current with multiplicity $p$ is obtained by the sum of the expressions
associated with all the $p+1$ cubic graphs with one leg off-shell, whose total number is given
by the Catalan number $C_{p-1}$. It is convenient to recall that the Catalan number $C_{p-1}$
represents the number of different ways that $p$ factors can be bracketed and each possibility has a
direct representation in terms of cubic graphs.
To each graph a BRST block $K_{[[ \ldots, \ldots], \ldots]}$ is assigned with the corresponding
bracketing (which reflects the vertex structure). In addition, an inverse Mandelstam invariant should be multiplied for each
non-external edge.

The two possibilities of bracketing three external legs, namely
$[[12]3]$ and  $[1[23]]$, give rise to the expression for ${\cal K}_{123}$ under the mapping described
above, see \figMthree. Similarly, the five different bracketing possibilities of four external legs
\eqn\BGfourbrackets{
[[[12]3]4],\;[[1[23]]4],\;[[12][34]],\;[1[2[34]]],\;[1[[23]4]]
}
and their corresponding mapping in terms of cubic graphs and BRST blocks leading to the expression
${\cal K}_{1234}$ were depicted in \BGfour.
Higher-multiplicity examples are similarly handled.

\ifig\figMthree{A diagrammatic derivation of the Berends--Giele current ${\cal K}_{123}$. The two cubic graphs
correspond to the two possibilities of bracketing three external legs, $[[12]3],[1[23]]$ and give rise
to the expression ${\cal K}_{123} = {K_{123}\over s_{12}s_{123}} + {K_{321}\over s_{23}s_{123}}$ under the mapping
described below together with the conventions \nested.}
{\epsfxsize=0.40\hsize\epsfbox{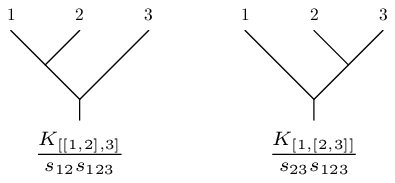}}


\subsec Different superfield representations versus Lie symmetries

The definition of the hatted BRST blocks at multiplicity $p$ has an explicit
antisymmetrization of the form $12 \ldots p-1 \leftrightarrow p$, where $p$ is a single-particle label. As discussed
above, the resulting BRST block is represented by a iterated bracket where the second slot of the outer bracket is a single-particle label.
This motivates to check the outcome of a more general hatted superfield definition featuring a multiparticle label instead of $p$.
As the brief discussion below
suggests, the result is compatible with a linear combination of the ``standard'' BRST blocks following from the
iterated bracket notation.

To see this, consider a rank-four hatted BRST block
with the symmetry structure $(12\leftrightarrow 34)$ instead of $(123\leftrightarrow 4)$ as in \Ahatfour. For example,
\eqn\NewAhatfour{
\hat A^{[[1,2],[3,4]]}_\a = - \half\Bigl[  A^{12}_\a (k^{12}\cdot  A^{34}) +  A^{12}_m (\g^m  W^{34})_\a -\(12\leftrightarrow 34\)\Bigr]\,.
}
It is not difficult to show that $\hat V_{[[1,2],[3,4]]}\equiv \l^\a \hat A^{[[1,2],[3,4]]}_\a$ satisfies
\eqn\QNewAhatfour{
Q\hat V_{[[1,2],[3,4]]} =
(k^1\cdot k^2)\bigl[ V_2\hat V_{341} - (1\leftrightarrow 2)\bigr]
+ (k^3\cdot k^4)\bigl[ V_3\hat V_{124} - (3\leftrightarrow 4)\bigr] 
+ (k^{12}\cdot k^{34}) V_{12}V_{34}\,.
}
where the equation of motion for $D_{(\a}\hat A_{\b)}$ was contracted with $\l^\a\l^\b$ for the sake of simplicity.
Therefore the redefinition
\eqn\NewAhatredef{
V_{[[1,2],[3,4]]} \equiv \hat V_{[[1,2],[3,4]]} + (k^1\cdot k^2)\bigl[V_2 H_{341} - (1\leftrightarrow 2)\bigr]
+ (k^3\cdot k^4)\bigl[V_3 H_{124} - (3\leftrightarrow 4)\bigr]
}
satisfies
\eqn\QNewAhatfour{
Q V_{[[1,2],[3,4]]} = QV_{1234} - QV_{1243}\,.
}
This is compatible with the expectation from the bracket notation since
$V_{[[1,2],[3,4]]} = V_{1234} - V_{1243}$, see \exampLie.

\appendix{B}{BCJ relations and one-loop scalar cohomology elements}
\applab\CAYMapp

\noindent
The scalar cohomology elements $C_{1|A,B,C}$ constructed in section 5.3 were argued in \oneloopbb\ to be linear
combinations of SYM tree-level amplitudes multiplied by quadratic polynomials of Mandelstam invariants. Momentum
conservation as well as BCJ and KK relations among color ordered SYM amplitudes $A^{\rm YM}(\ldots)$
\refs{\BCJ,\KKref} lead to a multitude of different such representations for $C_{1|A,B,C}$. In the following, we
provide convenient representations at all multiplicities\foot{The explicit representation given at multiplicity five
in \oneloopbb\ fails to satisfy the above criterion of having local Mandelstam coefficients along with $A^{\rm
YM}(\ldots)$. The six-point representation was given only indirectly as an expansion in terms of $A^{F^4}$, which
represent the $\a'^2$ corrections of the string tree-level amplitudes.} in the sense that the total number of terms is
systematically reduced and inverse powers of Mandelstam invariants are avoided. As we shall see, these $A^{\rm YM}$
representations of $C_{1|A,B,C}$ are intriguingly related to BCJ relations among tree-level
amplitudes.

\subsec A shuffle formula for BCJ relations

Let us first define an operation $S[A,B]$ which concatenates two multiparticle labels $A$ and $B$ with
Berends--Giele symmetries (see section 4.1) into
one such set,
\eqn\QEone{
M_{S[A,B]} \equiv \sum_{i=1}^{|A|} \sum_{j=1}^{|B|} (-1)^{i-j+|A|-1}
s_{a_i b_j} M_{(a_1 a_2\ldots a_{i-1} \shuffle a_{|A|} a_{|A|-1}\ldots a_{i+1})a_ib_j
(b_{j-1}\ldots b_2 b_1 \shuffle b_{j+1} \ldots b_{|B|})} .
}
One can interpret $M_{S[A,B]}$ in \QEone\ as attaching two Berends--Giele currents $M_A$ and $M_B$ to a cubic vertex
and expressing the resulting diagram in terms of $M_C$ at overall multiplicity $|C|=|A|+|B|$, see \figseven. For example,
\eqnn\QEoneEx
$$\eqalignno{
M_{S[1,2]} &= s_{12} M_{12}&\QEoneEx\cr
M_{S[1,23]} &= s_{12} M_{123} - s_{13}M_{132}\cr
M_{S[1,234]} &= s_{12} M_{1234} - s_{13}(M_{1324} + M_{1342}) + s_{14}M_{1432}\cr
M_{S[12,34]} &= - s_{13} M_{2134} + s_{14}M_{2143} + s_{23}M_{1234} - s_{24}M_{1243}\,.
}$$

\ifig\figseven{Diagrammatic interpretation of $M_{S[A,B]}$.}
{\epsfxsize=0.70\hsize\epsfbox{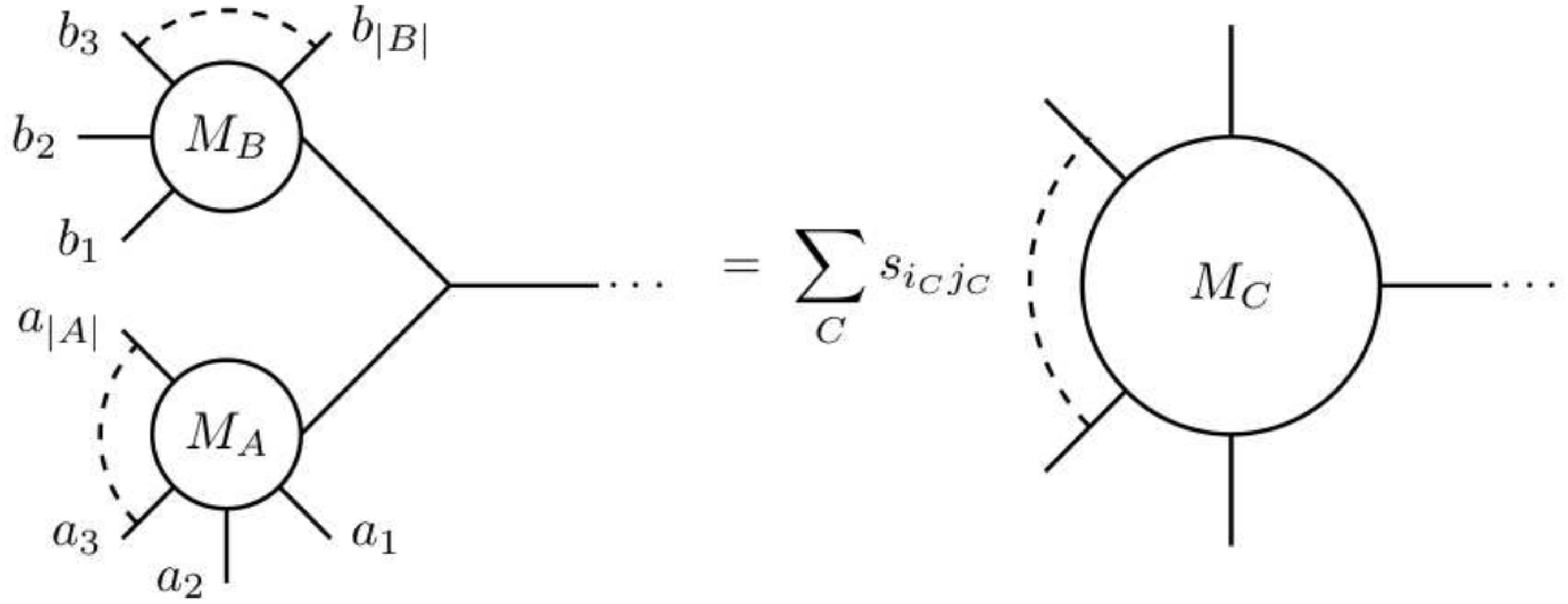}}


It turns out that the $S[A,B]$ product defined by \QEone\ can be used to generate BCJ relations among tree amplitudes \BCJ.
Recalling \nptMethod\ that SYM tree amplitudes are given by $A^{\rm YM}(1,2, \ldots, n) = \langle V_1 E_{23\ldots n} \rangle$,
BCJ relations among $A^{\rm YM}$ can be written as
\eqnn\altBCJ
$$\eqalignno{
&\langle V_1 E_{S[A,B]} \rangle = 0,\quad \forall \ A,B\,,  &\altBCJ
}$$
for example
\eqnn\BCJexex
$$\eqalignno{
0= \langle V_1 E_{S[2,34]}  \rangle &=s_{23}  A^{\rm YM}(1,2,3,4) - s_{24} A^{\rm YM}(1,2,4,3)  &\BCJexex\cr
0= \langle V_1E_{S[2,345]} \rangle &=s_{23} A^{\rm YM}(1,2,3,4,5) + s_{25} A^{\rm YM}(1,2,5,4,3)  \cr
&  - s_{24} (A^{\rm YM}(1,2,4,3,5)+A^{\rm YM}(1,2,4,5,3)) \cr
0= \langle V_1E_{S[23,45]}\rangle &= s_{34}  A^{\rm YM}(1,2,3,4,5)
- s_{35} A^{\rm YM}(1,2,3,5,4) \cr
& - s_{24}  A^{\rm YM}(1,3,2,4,5)+ s_{25} A^{\rm YM}(1,3,2,5,4)\,. \cr
0= \langle V_1E_{S[2,3456]} \rangle &= s_{23} A^{\rm YM}(1,2,3,4,5,6) - s_{26} A^{\rm YM}(1,2,6,5,4,3) \cr
&   - s_{24} (A^{\rm YM}(1,2,4,3,5,6)+ A^{\rm YM}(1,2,4,5,3,6)+A^{\rm YM}(1,2,4,5,6,3)) \cr
&  + s_{25} (A^{\rm YM}(1,2,5,6,4,3)+ A^{\rm YM}(1,2,5,4,6,3)+A^{\rm YM}(1,2,5,4,3,6))  \ .\cr
}$$
Similar formul{\ae} for BCJ relations using shuffle products can be found in\foot{We thank Henrik Johansson for pointing out
reference \ChenJXA.}
\refs{\BjerrumBohrRD, \StiebergerHQ, \ChenJXA}.
We have explicitly verified that \altBCJ\ holds up to multiplicity $|A|+|B|+1=7$ using the data from \WWW.

\subsec $\langle C_{1|A,B,C}\rangle$ from the BCJ shuffle formula

Since \altBCJ\ also holds for $A$ or $B$ of the form $S[C,D]$, we can iterate the product \QEone\ and generate
further vanishing identities for SYM subamplitudes from $E_{S[S[A,B],C]}$. Any partition of $A,B$ and $C$ leads
to an $A^{\rm YM}$ relation with local polynomials of degree two in Mandelstam invariants. The examples
\eqnn\BCJex
$$\eqalignno{
0= \langle V_1&E_{S[S[2,3],4]}  \rangle = s_{23} s_{34} A^{\rm YM}(1,2,3,4)
       - s_{23} s_{24} A^{\rm YM}(1,3,2,4) &\BCJex\cr
0= \langle V_1&E_{S[S[4,5],23]}  \rangle = 
- s_{34}s_{45} A^{\rm YM}(1,2,3,4,5)
+ s_{35}s_{45} A^{\rm YM}(1,2,3,5,4)\cr
&\quad{}\quad{}\quad{}\quad{}\quad{} \quad{}\quad{}+ s_{24}s_{45} A^{\rm YM}(1,3,2,4,5)
       - s_{25}s_{45} A^{\rm YM}(1,3,2,5,4) \cr
0= \langle V_1&E_{S[S[5,6],234]}  \rangle = 
       - s_{56} s_{45} A^{\rm YM}(1,2,3,4,5,6)
       + s_{56} s_{46} A^{\rm YM}(1,2,3,4,6,5)\cr
&\quad{}\quad{}\quad{}\quad{}\quad{}  \quad{}\quad{}     + s_{56} s_{35} A^{\rm YM}(1,2,4,3,5,6)
       - s_{56} s_{36} A^{\rm YM}(1,2,4,3,6,5)\cr
&\quad{} \quad{}\quad{}\quad{}\quad{}  \quad{}\quad{}    + s_{56} s_{35} A^{\rm YM}(1,4,2,3,5,6)
       - s_{56} s_{36} A^{\rm YM}(1,4,2,3,6,5)\cr
&\quad{}\quad{}\quad{}\quad{}\quad{}   \quad{}\quad{}    - s_{56} s_{25} A^{\rm YM}(1,4,3,2,5,6)
       + s_{56} s_{26} A^{\rm YM}(1,4,3,2,6,5) \cr
0= \langle V_1&E_{S[S[6,45],23]}  \rangle = 
       + s_{46} s_{34} A^{\rm YM}(1,2,3,4,5,6)
       + s_{56} s_{34} A^{\rm YM}(1,2,3,4,5,6)\cr
&\quad{}  \quad{}\quad{}\quad{}\quad{}  \quad{}\quad{}   + s_{46} s_{34} A^{\rm YM}(1,2,3,4,6,5)
       - s_{46} s_{35} A^{\rm YM}(1,2,3,5,4,6)\cr
&\quad{}  \quad{}\quad{}\quad{}\quad{}  \quad{}\quad{}   - s_{56} s_{35} A^{\rm YM}(1,2,3,5,4,6)
       - s_{56} s_{35} A^{\rm YM}(1,2,3,5,6,4)\cr
&\quad{}  \quad{}\quad{}\quad{}\quad{}  \quad{}\quad{}   - s_{46} s_{36} A^{\rm YM}(1,2,3,6,4,5)
       + s_{56} s_{36} A^{\rm YM}(1,2,3,6,5,4)\cr
&\quad{}  \quad{}\quad{}\quad{}\quad{}  \quad{}\quad{}   - s_{46} s_{24} A^{\rm YM}(1,3,2,4,5,6)
       - s_{56} s_{24} A^{\rm YM}(1,3,2,4,5,6)\cr
&\quad{}  \quad{}\quad{}\quad{}\quad{}  \quad{}\quad{}   - s_{46} s_{24} A^{\rm YM}(1,3,2,4,6,5)
       + s_{46} s_{25} A^{\rm YM}(1,3,2,5,4,6)\cr
&\quad{}  \quad{}\quad{}\quad{}\quad{}  \quad{}\quad{}   + s_{56} s_{25} A^{\rm YM}(1,3,2,5,4,6)
       + s_{56} s_{25} A^{\rm YM}(1,3,2,5,6,4)\cr
&\quad{}   \quad{}\quad{}\quad{}\quad{}  \quad{}\quad{}  + s_{46} s_{26} A^{\rm YM}(1,3,2,6,4,5)
       - s_{56} s_{26} A^{\rm YM}(1,3,2,6,5,4) 
}$$
can be checked to be a consequence of the BCJ relations \BCJ. Note that $E_{S[S[A,B],C]}$ in the five-point example is
chosen as $(A,B,C=4,5,23)$ rather than $(A,B,C=23,4,5)$ in order to minimize the number of terms.

The motivation to delve on the redundant BCJ relations \BCJex\ in addition to \BCJexex\ stems from their intriguing
connection with the $A^{\rm YM}$ representation of the scalar cohomology elements $C_{1|A,B,C}$. Up to six--points, we
have
\eqnn\CsToAYM
$$\eqalignno{
- \langle C_{1|2,3,4}\rangle & = - s_{24} s_{23} A^{\rm YM}(1,3,2,4) &\CsToAYM\cr
- \langle C_{1|23,4,5}\rangle & =
	- s_{45} s_{34} A^{\rm YM}(1,2,3,4,5)
        + s_{45} s_{24} A^{\rm YM}(1,3,2,4,5)\cr
-\langle C_{1|234,5,6}\rangle & =
       - s_{56} s_{45} A^{\rm YM}(1,2,3,4,5,6)
       + s_{56} s_{35} A^{\rm YM}(1,2,4,3,5,6)\cr
&\quad{}       + s_{56} s_{35} A^{\rm YM}(1,4,2,3,5,6)
       - s_{56} s_{25} A^{\rm YM}(1,4,3,2,5,6)\cr
-\langle C_{1|23,45,6}\rangle & =
       - s_{46} s_{36} A^{\rm YM}(1,2,3,6,4,5)
       + s_{56} s_{36} A^{\rm YM}(1,2,3,6,5,4)\cr
&\quad{}       + s_{46} s_{26} A^{\rm YM}(1,3,2,6,4,5)
       - s_{56} s_{26} A^{\rm YM}(1,3,2,6,5,4)\ ,\cr
}$$
and we observe that the expressions on the right hand side can be found by systematically deleting subsets of the
terms in \BCJex: Only those terms in $\langle V_1 E_{S[S[A,B],C]} \rangle$ are kept where the Mandelstam bilinear
takes the form $s_{ab}s_{ac}$ with  $a\in A$, $b\in B$ and $c\in C$. The following algorithm
allows to translate any $\langle C_{1|A,B,C}\rangle$ into SYM trees:
\medskip
\item{1.} Reorder the labels $A,B$ and $C$ such that $|A|\le|B|\le|C|$.
\item{2.} Apply the formula \QEone\ recursively to evaluate
$E_{S[S[A,B],C]}$.
\item{3.} Substitute $E_{\s_2 \ldots \s_{n}} \rightarrow A^{\rm YM}(1,\s_2, \ldots,\s_{n})$.
\item{4.} Keep only the terms containing Mandelstams with labels distributed as in $s_{ab}s_{ac}$, with
single-particle labels $a\in A$, $b\in B$ and $c\in C$. Delete terms of the form $s_{ab}s_{bc}$.
\item{5.} The result is $- \langle C_{1|A,B,C}\rangle$.
\medskip\noindent
We have explicitly checked with the data available from \WWW\ that the algorithm above is correct
for all scalar cohomology elements up to multiplicity $|A|+|B|+|C|+1=7$. For example, it leads to
\eqnn\moreCsToAYM
$$\eqalignno{
- \langle C_{1|2345,6,7}\rangle & =
       - s_{67} s_{56} A^{\rm YM}(1,2,3,4,5,6,7)
       + s_{67} s_{46} A^{\rm YM}(1,2,3,5,4,6,7) &\moreCsToAYM \cr
&\quad{}       + s_{67} s_{46} A^{\rm YM}(1,2,5,3,4,6,7)
       - s_{67} s_{36} A^{\rm YM}(1,2,5,4,3,6,7)\cr
&\quad{}       + s_{67} s_{46} A^{\rm YM}(1,5,2,3,4,6,7)
       - s_{67} s_{36} A^{\rm YM}(1,5,2,4,3,6,7)\cr
&\quad{}       - s_{67} s_{36} A^{\rm YM}(1,5,4,2,3,6,7)
       + s_{67} s_{26} A^{\rm YM}(1,5,4,3,2,6,7)\cr
-\langle C_{1|234,56,7} \rangle& =
 	- s_{57} s_{47} A^{\rm YM}(1,2,3,4,7,5,6)
        + s_{67} s_{47} A^{\rm YM}(1,2,3,4,7,6,5)\cr
 &\quad{}       + s_{57} s_{37} A^{\rm YM}(1,2,4,3,7,5,6)
        - s_{67} s_{37} A^{\rm YM}(1,2,4,3,7,6,5)\cr
 &\quad{}       + s_{57} s_{37} A^{\rm YM}(1,4,2,3,7,5,6)
        - s_{67} s_{37} A^{\rm YM}(1,4,2,3,7,6,5)\cr
 &\quad{}       - s_{57} s_{27} A^{\rm YM}(1,4,3,2,7,5,6)
        + s_{67} s_{27} A^{\rm YM}(1,4,3,2,7,6,5)\cr
 %
}$$
and a slightly longer 32-term representation of $\langle C_{1|23,45,67} \rangle$ which is commented out in the \TeX\ source.

It will be interesting to understand the origin of the intriguing patterns described in this
Appendix. They hint a deeper connection between the fusion of Berends--Giele currents via \QEone\ (see \figseven\ for a
diagrammatic interpretation), general BCJ relations \refs{\BCJ,\ChenJXA} and the scalar cohomology elements $\langle C_{1|A,B,C} \rangle$ generating
the non-anomalous kinematics in one-loop amplitudes of the open superstring \oneloopbb.

\appendix{C}{The explicit expression for $H_{1234}$}
\applab\Hfourapp

\noindent The Lie symmetry of rank-four BRST blocks is restored by the redefinition \redefAmfour\ with the following
expression for $H_{1234}$:
\eqn\Hfoursol{
4H_{1234} = H^{(a)}_{1234} - H^{(a)}_{1243} + H^{(a)}_{3412} - H^{(a)}_{3421}
}
By construction, it is the $\lie_4$ image of a more elementary expression
\eqnn\HfoursolO
$$\eqalignno{
H^{(a)}_{1234} &=
{1\over 4}(A^{12}\cdot A^{34})
+{1 \over 6}\, (A^{12}\cdot A^{3}) (k^3 \cdot A^4)
 -  {1 \over 3}\, (A^{12}\cdot A^{3}) (k^{12} \cdot A^4)
 +  {1 \over 2}\, (A^{123}\cdot A^{4})\cr
&{} +  {1 \over 2}\, A^{12}_{m} A^{3}_{n} F^{4}_{mn}
 +  {1 \over 6}\, (A^{1}\cdot A^{23}) (k^{123} \cdot A^4)
 -  {1 \over 6}\, (A^{2}\cdot A^{13}) (k^{123} \cdot A^4)\cr
&{} + {1\over 4}\bigl(H^{(b)}_{1234} + H^{(b)}_{3412}
 + H^{(b)}_{1423}
 + H^{(b)}_{2314}
 + H^{(b)}_{3124}
 + H^{(b)}_{2431}\bigr) &\HfoursolO
}$$
with
\eqnn\HfoursolTwo
$$\eqalignno{
H^{(b)}_{1234} &=   {1 \over 6}\, (A^1\cdot A^2) (k^4 \cdot A^3) ((k^1 \cdot A^4) - (k^2 \cdot A^4))&\HfoursolTwo\cr
&{} -  {1 \over 6}\, (A^1\cdot A^2) (k^3 \cdot A^4) ((k^1 \cdot A^3) - (k^2 \cdot A^3)) \cr
&{} +  {1 \over 3}\, (A^1\cdot A^2)
\bigl[(k^2 \cdot A^3) (k^1 \cdot A^4) - (k^1 \cdot A^3) (k^2 \cdot A^4)\bigr] \ .
}$$

\listrefs

\bye

%% file: harvmacM.tex

%
\def\unredoffs{}
\tolerance=1000\hfuzz=2pt
\catcode`\@=11 
\ifx\hyperdef\UNd@FiNeD\def\hyperdef#1#2#3#4{#4}\def\hyperref#1#2#3#4{#4}\def\href#1#2{#2}\fi
\magnification=1200\unredoffs\baselineskip=16pt plus 2pt minus 1pt
\def\Date#1{\vfill\leftline{#1}\tenpoint\supereject%
\footline={\hss\tenrm\hyperdef\hypernoname{page}\folio\folio\hss}}%

{\count255=\time\divide\count255 by 60 \xdef\hourmin{\number\count255}
 \multiply\count255 by-60\advance\count255 by\time
 \xdef\hourmin{\hourmin:\ifnum\count255<10 0\fi\the\count255}
}
\def\date{\number\day.\number\month.\number\year\ at \hourmin}


\def\nolabels{\def\wrlabeL##1{}\def\eqlabeL##1{}\def\reflabeL##1{}}
\def\writelabels{\def\wrlabeL##1{\leavevmode\vadjust{\rlap{\smash%
{\line{{\escapechar=` \hfill\rlap{\sevenrm\hskip.03in\string##1}}}}}}}%
\def\eqlabeL##1{{\escapechar-1\rlap{\sevenrm\hskip.05in\string##1}}}%
\def\reflabeL##1{\noexpand\llap{\noexpand\sevenrm\string\string\string##1}}}
\nolabels

\global\newcount\secno \global\secno=0
\global\newcount\meqno \global\meqno=1
\def\s@csym{}

\def\newsec#1\par{\global\advance\secno by1%
{\toks0{#1}\message{(\the\secno. \the\toks0)}}%
\global\subsecno=0\eqnres@t\let\s@csym\secsym\xdef\secn@m{\the\secno}\noindent
{\bf\hyperdef\hypernoname{section}{\the\secno}{\the\secno.} #1}%
\writetoca{{\string\hyperref{}{section}{\the\secno}{\bf \the\secno\quad}} {\bf #1}}%
\par\nobreak\medskip\nobreak\noindent\ignorespaces}
\def\eqnres@t{\xdef\secsym{\the\secno.}\global\meqno=1\bigbreak\bigskip}
\def\sequentialequations{\def\eqnres@t{\bigbreak}}\xdef\secsym{}

\global\newcount\subsecno \global\subsecno=0
\def\subsec#1\par{\global\advance\subsecno by1%
{\toks0{#1}\message{(\s@csym\the\subsecno. \the\toks0)}}%
\global\subsubsecno=0%
\ifnum\lastpenalty>9000\else\bigbreak\fi
\noindent{\it\hyperdef\hypernoname{subsection}{\secn@m.\the\subsecno}%
{\secn@m.\the\subsecno.} #1}\writetoca{\string\hskip1.45cm
{\string\hyperref{}{subsection}{\secn@m.\the\subsecno}{\secn@m.\the\subsecno.}}
{#1}}\par\nobreak\medskip\nobreak\noindent\ignorespaces}

\def\appendix#1#2{\global\meqno=1\global\subsecno=0\xdef\secsym{\hbox{#1.}}%
\bigbreak\bigskip\noindent{\bf Appendix \hyperdef\hypernoname{appendix}{#1}%
{#1.} #2}{\toks0{(#1. #2)}\message{\the\toks0}}%
\xdef\s@csym{#1.}\xdef\secn@m{#1}%
\writetoca{{\string\hyperref{}{appendix}{#1}{\bf {#1}\quad}} {\bf #2}}%
\par\nobreak\medskip\nobreak}

%
\def\checkm@de#1#2{\ifmmode{\def\f@rst##1{##1}\hyperdef\hypernoname{equation}%
{#1}{#2}}\else\hyperref{}{equation}{#1}{#2}\fi}
\def\eqnn#1{\DefWarn#1\xdef #1{(\noexpand\relax\noexpand\checkm@de%
{\s@csym\the\meqno}{\secsym\the\meqno})}%
\wrlabeL#1\writedef{#1\leftbracket#1}\global\advance\meqno by1}
\def\f@rst#1{\c@t#1a\em@ark}\def\c@t#1#2\em@ark{#1}
\def\eqna#1{\DefWarn#1\wrlabeL{#1$\{\}$}%
\xdef #1##1{(\noexpand\relax\noexpand\checkm@de%
{\s@csym\the\meqno\noexpand\f@rst{##1}1}{\hbox{$\secsym\the\meqno##1$}})}
\writedef{#1\numbersign1\leftbracket#1{\numbersign1}}\global\advance\meqno by1}
\def\eqn#1#2{\DefWarn#1%
\xdef #1{(\noexpand\hyperref{}{equation}{\s@csym\the\meqno}%
{\secsym\the\meqno})}$$#2\eqno(\hyperdef\hypernoname{equation}%
{\s@csym\the\meqno}{\secsym\the\meqno})\eqlabeL#1$$%
\writedef{#1\leftbracket#1}\global\advance\meqno by1}
\def\xeqn{\expandafter\xe@n}\def\xe@n(#1){#1}
\def\xeqna#1{\expandafter\xe@n#1}
\def\eqns#1{(\e@ns #1{\hbox{}})}
\def\e@ns#1{\ifx\UNd@FiNeD#1\message{eqnlabel \string#1 is undefined.}%
\xdef#1{(?.?)}\fi{\let\hyperref=\relax\xdef\next{#1}}%
\ifx\next\em@rk\def\next{}\else%
\ifx\next#1\xeqn#1\else\def\n@xt{#1}\ifx\n@xt\next#1\else\xeqna#1\fi
\fi\let\next=\e@ns\fi\next}

\def\DefWarn#1{\ifx\UNd@FiNeD#1\else
\immediate\write16{*** WARNING: the label \string#1 is already defined ***}\fi}
%
\newskip\footskip\footskip14pt plus 1pt minus 1pt 
\def\footnotefont{\ninepoint}\def\f@t#1{\footnotefont #1\@foot}
\def\f@@t{\baselineskip\footskip\bgroup\footnotefont\aftergroup\@foot\let\next}
\setbox\strutbox=\hbox{\vrule height9.5pt depth4.5pt width0pt}
\global\newcount\ftno \global\ftno=0
\def\foot{\global\advance\ftno by1\def\foot@rg{\hyperref{}{footnote}%
{\the\ftno}{\the\ftno}\xdef\foot@rg{\noexpand\hyperdef\noexpand\hypernoname%
{footnote}{\the\ftno}{\the\ftno}}}\footnote{$^{\foot@rg}$}}
%
%
%
\global\newcount\refno \global\refno=1
\newwrite\rfile
\def\ref{[\hyperref{}{reference}{\the\refno}{\the\refno}]\nref}
\def\nref#1{\DefWarn#1%
\xdef#1{[\noexpand\hyperref{}{reference}{\the\refno}{\the\refno}]}%
\writedef{#1\leftbracket#1}%
\ifnum\refno=1\immediate\openout\rfile=\jobname.refs\fi
\chardef\wfile=\rfile\immediate\write\rfile{\noexpand\item{[\noexpand\hyperdef%
\noexpand\hypernoname{reference}{\the\refno}{\the\refno}]\ }%
\reflabeL{#1\hskip.31in}\pctsign}\global\advance\refno by1\findarg}
\def\findarg#1#{\begingroup\obeylines\newlinechar=`\^^M\pass@rg}
{\obeylines\gdef\pass@rg#1{\writ@line\relax #1^^M\hbox{}^^M}%
\gdef\writ@line#1^^M{\expandafter\toks0\expandafter{\striprel@x #1}%
\edef\next{\the\toks0}\ifx\next\em@rk\let\next=\endgroup\else\ifx\next\empty%
\else\immediate\write\wfile{\the\toks0}\fi\let\next=\writ@line\fi\next\relax}}
\def\striprel@x#1{} \def\em@rk{\hbox{}}
\def\lref{\begingroup\obeylines\lr@f}
\def\lr@f#1#2{\DefWarn#1\gdef#1{\let#1=\UNd@FiNeD\ref#1{#2}}\endgroup\unskip}
\def\semi{;\hfil\break}
\def\addref#1{\immediate\write\rfile{\noexpand\item{}#1}} 
\def\listrefs{\vfill\supereject\immediate\closeout\rfile\writestoppt
\baselineskip=\footskip\centerline{{\bf References}}\bigskip{\parindent=20pt%
\frenchspacing\escapechar=` \input \jobname.refs\vfill\eject}\nonfrenchspacing}
\def\startrefs#1{\immediate\openout\rfile=\jobname.refs\refno=#1}
\def\xref{\expandafter\xr@f}\def\xr@f[#1]{#1}
\def\refs#1{\count255=1[\r@fs #1{\hbox{}}]}
\def\r@fs#1{\ifx\UNd@FiNeD#1\message{reflabel \string#1 is undefined.}%
\nref#1{need to supply reference \string#1.}\fi%
\vphantom{\hphantom{#1}}{\let\hyperref=\relax\xdef\next{#1}}%
\ifx\next\em@rk\def\next{}%
\else\ifx\next#1\ifodd\count255\relax\xref#1\count255=0\fi%
\else#1\count255=1\fi\let\next=\r@fs\fi\next}
%

%
\newwrite\ffile\global\newcount\figno \global\figno=1
\def\fig{fig.~\hyperref{}{figure}{\the\figno}{\the\figno}\nfig}
\def\nfig#1{\DefWarn#1%
\xdef#1{fig.~\noexpand\hyperref{}{figure}{\the\figno}{\the\figno}}%
\writedef{#1\leftbracket fig.\noexpand~\xfig#1}%
\ifnum\figno=1\immediate\openout\ffile=\jobname.figs\fi\chardef\wfile=\ffile%
{\let\hyperref=\relax
\immediate\write\ffile{\noexpand\medskip\noexpand\item{Fig.\ %
\noexpand\hyperdef\noexpand\hypernoname{figure}{\the\figno}{\the\figno}. }
\reflabeL{#1\hskip.55in}\pctsign}}\global\advance\figno by1\findarg}
\def\xfig{\expandafter\xf@g}\def\xf@g fig.\penalty\@M\ {}
\def\figs#1{figs.~\f@gs #1{\hbox{}}}
\def\f@gs#1{{\let\hyperref=\relax\xdef\next{#1}}\ifx\next\em@rk\def\next{}\else
\ifx\next#1\xfig #1\else#1\fi\let\next=\f@gs\fi\next}
%
\def\figin{\epsfcheck\figin}\def\figins{\epsfcheck\figins}
\def\epsfcheck{\ifx\epsfbox\UnDeFiNeD
\message{(NO epsf.tex, FIGURES WILL BE IGNORED)}
\gdef\figin##1{\vskip2in}\gdef\figins##1{\hskip.5in}
\else\message{(FIGURES WILL BE INCLUDED)}%
\gdef\figin##1{##1}\gdef\figins##1{##1}\fi}
\def\DefWarn#1{}
\def\figinsert{\goodbreak\topinsert}
\def\ifig#1#2#3{\DefWarn#1\xdef#1{fig.~\the\figno}
\writedef{#1\leftbracket fig.\noexpand~\the\figno}%
\figinsert\figin{\centerline{#3}}
\smallskip
\leftskip=20pt \rightskip=20pt
\baselineskip12pt\noindent
{{\bf Fig.~\the\figno}\ \ninepoint #2}
\medskip
\global\advance\figno by1\par\endinsert}
\newwrite\lfile
{\escapechar-1\xdef\pctsign{\string\%}\xdef\leftbracket{\string\{}
\xdef\rightbracket{\string\}}\xdef\numbersign{\string\#}}
\def\writedefs{\immediate\openout\lfile=label.defs \def\writedef##1{%
{\let\hyperref=\relax\let\hyperdef=\relax\let\hypernoname=\relax
 \immediate\write\lfile{\string\def\string##1\rightbracket}}}}%
\def\writestop{\def\writestoppt{\immediate\write\lfile{\string\pageno
 \the\pageno\string\startrefs\leftbracket\the\refno\rightbracket
 \string\def\string\secsym\leftbracket\secsym\rightbracket
 \string\secno\the\secno\string\meqno\the\meqno}\immediate\closeout\lfile}}
\def\writestoppt{}\def\writedef#1{}

\def\seclab#1{\DefWarn#1%
\xdef #1{\noexpand\hyperref{}{section}{\the\secno}{\the\secno}}%
\writedef{#1\leftbracket#1}\wrlabeL{#1=#1}}
\def\subseclab#1{\DefWarn#1%
\xdef #1{\noexpand\hyperref{}{subsection}{\the\secno.\the\subsecno}%
{\the\secno.\the\subsecno}}\writedef{#1\leftbracket#1}\wrlabeL{#1=#1}}
\def\applab#1{\DefWarn#1%
\xdef #1{\noexpand\hyperref{}{appendix}{\secn@m}{\secn@m}}%
\writedef{#1\leftbracket#1}\wrlabeL{#1=#1}}
\newwrite\tfile \def\writetoca#1{}
\def\leaderfill{\leaders\hbox to 1em{\hss.\hss}\hfill}
\def\writetoc{\immediate\openout\tfile=\jobname.toc
   \def\writetoca##1{{\edef\next{\write\tfile{\noindent ##1
   \string\leaderfill{
   \string\hyperref{}{page}{\noexpand\number\pageno}%
   {\noexpand\number\pageno}} \par}}\next}}
}
\newread\ch@ckfile
\def\listtoc{\immediate\closeout\tfile\immediate\openin\ch@ckfile=\jobname.toc
\ifeof\ch@ckfile\message{no file \jobname.toc, no table of contents this pass}%
\else\closein\ch@ckfile\centerline{\bf Contents}\nobreak\medskip%
{\baselineskip=16pt\footnotefont\parskip=0pt\catcode`\@=11\input\jobname.toc
\catcode`\@=12\bigbreak\bigskip}\fi}
\catcode`\@=12 
\def\tenpoint{\def\rm{\fam0\tenrm}
\textfont0=\tenrm \scriptfont0=\sevenrm \scriptscriptfont0=\fiverm
\textfont1=\teni  \scriptfont1=\seveni  \scriptscriptfont1=\fivei
\textfont2=\tensy \scriptfont2=\sevensy \scriptscriptfont2=\fivesy
\textfont\itfam=\tenit \def\it{\fam\itfam\tenit}\def\footnotefont{\ninepoint}%
\textfont\bffam=\tenbf \def\bf{\fam\bffam\tenbf}\def\sl{\fam\slfam\tensl}\rm}
\font\ninerm=cmr9 \font\sixrm=cmr6 \font\ninei=cmmi9 \font\sixi=cmmi6
\font\ninesy=cmsy9 \font\sixsy=cmsy6 \font\ninebf=cmbx9
\font\nineit=cmti9 \font\ninesl=cmsl9 \skewchar\ninei='177
\skewchar\sixi='177 \skewchar\ninesy='60 \skewchar\sixsy='60
\def\ninepoint{\def\rm{\fam0\ninerm}
\textfont0=\ninerm \scriptfont0=\sixrm \scriptscriptfont0=\fiverm
\textfont1=\ninei \scriptfont1=\sixi \scriptscriptfont1=\fivei
\textfont2=\ninesy \scriptfont2=\sixsy \scriptscriptfont2=\fivesy
\textfont\itfam=\ninei \def\it{\fam\itfam\nineit}\def\sl{\fam\slfam\ninesl}%
\textfont\bffam=\ninebf \def\bf{\fam\bffam\ninebf}\rm}
%
\hyphenation{anom-aly anom-alies coun-ter-term coun-ter-terms}

\global\newcount\subsubsecno \global\subsubsecno=0
\def\subsubsec#1\par{\global\advance\subsubsecno by1%
{\toks0{#1}\message{(\the\secno\the\subsecno\the\subsubsecno. \the\toks0)}}%
\ifnum\lastpenalty>9000\else\bigbreak\fi
\noindent{\it\hyperdef\hypernoname{subsubsection}{\the\secno.\the\subsecno\the\subsubsecno}%
{\the\secno.\the\subsecno.\the\subsubsecno.} #1}
\par\nobreak\medskip\nobreak\noindent\ignorespaces}

\def\DefWarn#1{}
\def\tikzcaption#1#2{\DefWarn#1\xdef#1{Fig.~\the\figno}
\writedef{#1\leftbracket Fig.\noexpand~\the\figno}%
{
\smallskip
\leftskip=20pt \rightskip=20pt \baselineskip12pt\noindent
{{\bf Fig.~\the\figno}\ \ninepoint #2}
\bigskip
\global\advance\figno by1 \par}}

\def\ntoalpha#1{%
\ifcase#1%
@%
\or A\or B\or C\or D\or E\or F\or G\or H\or I
\fi
}

\global\newcount\appno \global\appno=1
\def\applab#1{\xdef #1{\ntoalpha\appno}\writedef{#1\leftbracket#1}\wrlabeL{#1=#1}
\global\advance\appno by1}

\def\preprint#1 #2\par{\rightline{\vbox{\baselineskip12pt\hbox{#1}\hbox{#2}}}\vskip2cm}
%
\def\title#1\par{\centerline{\bf #1}\nopagenumbers\pageno=0}
\def\author#1\par{\bigskip\bigskip\centerline{#1}}

\newcount\addressno

\def\email#1#2{\unskip$^#1$\footnote{\null}{\kern-\parindent \llap{$^#1$\hskip1pt}email: #2}}

\def\startcenter{%
  \par
  \begingroup
  \leftskip=0pt plus 1fil
  \rightskip=\leftskip
  \parindent=0pt
  \parfillskip=0pt
}
\def\stopcenter{\endgroup}

\def\address{\bigskip%
  \ifnum\the\addressno=0\else\stopcenter\endgroup\fi
  \advance\addressno by 1%
  \begingroup
  \startcenter
  \it
  \obeylines
  \addressAux
}
\def\addressAux#1{#1}

\def\abstract{\stopcenter\endgroup\bigskip\bigskip\noindent}

\def\Dsl{\,\raise.15ex\hbox{/}\mkern-13.5mu D} 
\def\dsl{\raise.15ex\hbox{/}\kern-.57em\partial}
 
\def\boxeqn#1{\vcenter{\vbox{\hrule\hbox{\vrule\kern3pt\vbox{\kern3pt
	\hbox{${\displaystyle #1}$}\kern3pt}\kern3pt\vrule}\hrule}}}

\def\lie{\hbox{\it\$}} 

\def\a{\alpha}
\def\b{{\beta}}
\def\g{{\gamma}}
\def\d{{\delta}}

\def\l{\lambda}

\def\s{{\sigma}}
\def\t{{\theta}}

\def\half{{1\over 2}}
\def\p{{\partial}}

\def\({\left(}
\def\){\right)}
\def\cF{{\cal F}}
\def\cW{{\cal W}}

\def\cA{{\cal A}}
\def\cV{{\cal V}}

\font\tenshuffle=shuffle10 \font\sevenshuffle=shuffle7 \font\fiveshuffle=shuffle7 at 5pt
\def\shuffle{{%
\def\Dshuffle{\mathbin{\hbox{\tenshuffle\char'001}}}%
\def\Sshuffle{\mathbin{\hbox{\sevenshuffle\char'001}}}%
\def\SSshuffle{\mathbin{\hbox{\fiveshuffle\char'001}}}%
\mathchoice{\Dshuffle}{\Dshuffle}{\Sshuffle}{\SSshuffle}}}


\def\qed{\hbox{\hskip 3pt
\vbox{\hrule\hbox to 7pt{\vrule height 7pt\hfill\vrule}
\hrule}}\hskip3pt}

\overfullrule=0pt\relax

\frenchspacing

\newread\instream \openin\instream= label.defs
\ifeof\instream \message{No labels in advance yet. Wait till next pass.}
\else \closein\instream \input label.defs
\fi
\writedefs

\def\arXiv:#1].{\hepthStrip#1 \nil}
\def\hepthStrip#1 #2\nil{\href{http://arxiv.org/abs/#1}{arXiv:#1 #2\unskip}].}